\documentclass[preprint,superscriptaddress,floatfix]{revtex4-2}

\usepackage{graphicx,footmisc}
\usepackage[caption=false]{subfig}
\usepackage{color}
\usepackage{natbib,url}
\usepackage{amsmath}
\usepackage{float}
\usepackage[normalem]{ulem}
\usepackage{longtable}
\usepackage{soul}

\newcommand*{\be}{\begin{equation}}
\newcommand*{\ee}{\end{equation}}

\def\begineq{\begin{equation}}
\def\endeq{\end{equation}}

\def\begineqn{\begin{equation*}}
\def\endeqn{\end{equation*}}

\def\beginar{\begin{eqnarray}}
\def\endar{\end{eqnarray}}
\def\beginarn{\begin{eqnarray*}}
\def\endarn{\end{eqnarray*}}

\def\lb{\left ( }
\def\rb{\right ) }
\def\lsq{\left [ }
\def\rsq{\right ] }

\def\ub{\mathbf{u}}

\def\Bb{\mathbf{B}}

\def\bb{\mathbf{b}}

\def\dsx{{\partial_x}}

\def\dst{{\partial_t}}

\def\dsz{{\partial_z}}

\def\hz{{\bf\widehat z}}

\def\hx{{\bf\widehat x}}

\def\lap{{\nabla^2}}

\def\kb{\mathbf{k}}

\begin{document}

\title{A numerical investigation of quasi-static magnetoconvection with an imposed horizontal magnetic field}

\author{Michael A. Calkins}%
\affiliation{ 
Department of Physics, University of Colorado, Boulder, Colorado 80309, USA
}%
\author{Talal AlRefae}
\affiliation{ 
Department of Physics, University of Colorado, Boulder, Colorado 80309, USA
}%
\author{Angel Hernandez}%
\affiliation{ 
Department of Physics, University of Colorado, Boulder, Colorado 80309, USA
}%
\author{Ming Yan}
\affiliation{ 
Department of Physics, University of Colorado, Boulder, Colorado 80309, USA
}%
\author{Stefano Maffei}
\affiliation{ 
Institute of Geophysics, ETH Z\"urich, Sonneggstrasse 5, 8092 Z\"urich, Switzerland
}%

\begin{abstract}



Quasi-static Rayleigh-B\'enard convection with an imposed horizontal magnetic field is investigated numerically for Chandrasekhar numbers up to $Q=10^6$ with stress free boundary conditions. 
Both $Q$ and the Rayleigh number ($Ra$) are varied to identify the various dynamical regimes that are present in this system.
We find three primary regimes:
(I) a two-dimensional (2D) regime in which the axes of the convection rolls are oriented parallel to the imposed magnetic field; (II) an anisotropic three-dimensional (3D) regime; and (III) a mean flow regime characterized by a large scale horizontal flow directed transverse to the imposed magnetic field. 
The transition to 3D dynamics is preceded by a series of 2D transitions in which the number of convective rolls decreases as $Ra$ is increased. For sufficiently large $Q$, there is an eventual transition to two rolls just prior to the 2D/3D transition. The 2D/3D transition occurs when inertial forces become comparable to the Lorentz force, i.e. when $\sqrt{Q}/Re = O(1)$; 2D, magnetically constrained states persist when $\sqrt{Q}/Re \gtrsim O(1)$. 
Within the 2D regime we find heat and momentum transport scalings that are consistent with the hydrodynamic asymptotic predictions of \citeauthor{gC09} [Phys. Fluids \textbf{21}, 083603 (2009)]: the Nusselt number ($Nu$) and Reynolds number ($Re$) scale as $Nu \sim Ra^{1/3}$ and $Re \sim Ra^{2/3}$, respectively. For $Q=10^6$, we find that the scaling behavior of $Nu$ and $Re$ breaks down at large values of $Ra$ due to a sequence of bifurcations and the eventual manifestation of mean flows.


\end{abstract}

\maketitle

\section{Introduction}

Convection-driven flows are important in a variety of engineering and natural systems. In the context of geophysical and astrophysical fluid systems, magnetic fields can play a significant role in the convection dynamics, and lead to dynamical states that would otherwise be absent \citep{jmA01,uB02,eK15,yT16,xY18,tV18,wL18,mY19,zL19,tZ20}. In particular, the motion of electrically conducting fluids in the presence of magnetic fields leads to conversions between magnetic and kinetic energy and an additional source of dissipation that results in anisotropy in the flow field. In convection-driven flows these effects can lead to heat and momentum transport that are different than the corresponding hydrodynamic case. Understanding how the magnetic field influences this transport is one of the primary goals of this study.

Magnetoconvection (MC) typically refers to convection of an electrically conducting fluid in the presence of an externally-controlled magnetic field $\mathbf{B}_0$ \citep{nW14}. Here we focus on the Rayleigh-B\'enard configuration with two parallel plates confining a Boussinesq fluid layer of depth $H$. The dynamics of MC depend upon both the magnitude and direction of the imposed field. MC has been investigated both experimentally and numerically for a variety of magnetic field directions (as measured relative to the direction of gravity), including vertical (VMC) \citep{jmA01,uB01,pM95,pM99,sC00,mY19}, horizontal (HMC) \citep{uB02,tV18} and tilted (TMC) \citep{nH96,jN22}. The linear behavior associated with these various field orientations provides an initial guide for understanding the resulting nonlinear behavior. The strength of the buoyancy force is quantified by the Rayleigh number,
\be
Ra = \frac{g \alpha \Delta T H^3}{\nu \kappa} ,
\ee
where $g$ is the (constant) gravitational acceleration, $\alpha$ is the thermal expansion coefficient, $\Delta T$ is the temperature difference between the two plates, $\nu$ is the kinematic viscosity and $\kappa$ is the thermal diffusivity. 
For VMC, the system is stabilized and therefore requires a larger critical Rayleigh number, $Ra_c$, to induce convection as the strength of the imposed field is increased \citep[e.g.][]{sC61}. Vertical motions are preferred in this case, and anisotropic, magnetically-aligned structures persist so long as the Lorentz force remains dominant relative to inertia \citep{mY19,tZ20}. For HMC, the most unstable mode consists of two-dimensional (2D) rolls oriented with their axes parallel to $\Bb_0$ \citep{sC61}. The resulting flows do not induce magnetic field, and therefore both the critical Rayleigh number and critical horizontal wavenumber are equal to their hydrodynamic values, $Ra_c = 27 \pi^4/4   \approx 658$ and $k_c \approx 2.22$, respectively, for the stress-free boundary conditions used in the present study. For a sufficiently strong horizontal magnetic field, previous work has shown that 2D rolls can persist at increasingly larger Rayleigh numbers as the magnitude of the field is increased, but an eventual transition to three-dimensional (3D) states is observed \citep{uB02,aO03,yT16,tV18,jY21}. \textcolor{black}{A second goal of the present study is to determine the parameter values at which this transition takes place, and to explore the dynamics of these 3D states.}


Global heat and momentum transport in convection are  characterized by the Nusselt number, $Nu$, and the Reynolds number, $Re = U H/\nu$ (where $U$ is a typical flow speed), respectively. The Nusselt number is the ratio of total heat transport (convective and conductive) to conductive heat transport in the absence of convection. Experiments and simulations of Rayleigh-B\'enard convection (RBC) show that both of these quantities tend to follow power-law scaling behavior for sufficiently large values of $Ra$ \citep{gA09,jC15}. The Nusselt number is observed to scale as $Nu \sim Ra^{\beta}$, where $\beta$ is a constant. Theoretical considerations for RBC find heat transport scalings of either $\beta=1/3$ or $\beta=1/2$; the former scaling is independent of the height of the fluid layer and heat transport is therefore controlled by conduction across the thermal boundary layers adjacent to the boundaries \citep{wM54}, whereas the latter scaling  is independent of diffusion coefficients and therefore considered to be representative of a state in which the entire fluid layer is turbulent \citep{rK62}. Simulations in triply-periodic domains have observed the $\beta=1/2$ exponent \citep{dL03}, though laboratory experiments and simulations with thermal (and momentum) boundary layers find exponents smaller than this upper bound. At moderate values of $Ra$, the $Nu \sim Ra^{2/7}$ scaling is well-documented \citep{bC89}. Experiments at larger values of $Ra$ have found $Nu \sim Ra^{0.322}$ \citep{jC15}, and recent two-dimensional numerical simulations find a $Nu \sim Ra^{0.35}$ scaling for $Ra \gtrsim 10^{13}$ in the hypothesized `ultimate' regime of fully-turbulent flow \citep{zhu2018transition}, though see \citep{cD19} for an alternative interpretation. 

The convective 	`free-fall' scaling for momentum transport, which can be obtained by  balancing nonlinear advection and inertia \citep[e.g.][]{jmA20}, leads to $Re \sim (Ra/Pr)^{1/2}$ \citep{rK62,eS65}, where $Pr = \nu / \kappa$ is the Prandtl number. This scaling is diffusion-free since both the thermal and the momentum diffusion coefficients cancel from the left and right hand sides of the relationship. The free-fall scaling fits data from triply periodic simulations well \citep{dL03}, and laboratory experiments and numerical simulations find scalings with a Rayleigh number exponent close to $1/2$ \citep{gA09,tV18b,mY19,tV21}. We note that for non-rotating convection-driven dynamos, heat and momentum transport seem to follow the non-magnetic scaling laws, even when ohmic dissipation and viscous dissipation are of comparable magnitude \citep{mY21}. 

One important goal for studies of steady-state 2D RBC is to identify flow morphologies and the associated length scales that maximize heat transport. For stress-free boundary conditions, Ref.~\citep{gC09} found that convective rolls with order unity aspect ratio maximize heat transport and follow a $Nu \sim Ra^{1/3}$ scaling law. These rolls are consistent with a $Re \sim Ra^{2/3}$ momentum transport \citep{bW20}. These coherent flow structures may persist within turbulent convection, and therefore may play an important role in controlling the transfer of heat and momentum \citep[e.g.][]{fW15,aP18,dK20}.

In comparison to RBC, heat and momentum transport has been studied less extensively in MC \citep[e.g.][]{jmA01,tZ16}. Recent work with a vertical magnetic field shows that both heat and momentum transport increase at a faster rate with increasing Rayleigh number \citep{mY19,tZ20}, though always at a value that is reduced relative to RBC. In contrast, in the case of a horizontal magnetic field, both laboratory and numerical experiments show that heat transport can be enhanced relative to RBC for certain regions of parameter space \citep{aO03,tV21b}. As for the case of 2D RBC, this enhancement is likely due to the presence of coherent flow structures that characterize certain flow regimes in MC with a horizontal field.

Strong horizontal mean flows can develop in 2D convection \citep{dG14} and 3D convection with a horizontal rotation vector \citep{von2015generation} if horizontally periodic boundary conditions and stress-free boundary conditions are used. These flows are characterized by a nearly linear vertical shear and can strongly influence the resulting convection, leading to time varying dynamics that can exhibit large amplitude changes in heat transport \citep{dG14,von2015generation,qW20}. The stability or existence of these mean flows depends on the particular combination of the non-dimensional parameters, but it seems that smaller domain aspect ratios favor their formation for a fixed value of the Rayleigh number \citep{qW20}.

In the present work we use direct numerical simulations to investigate MC in the presence of a horizontal magnetic field. A parameter survey is carried out to determine the heat and momentum transport scaling behavior. The resulting diagnostic quantities are used to characterize the different flow regimes, including the transition from 2D to 3D convection. 
The governing equations and numerical methods are discussed in section \ref{S:methods}, the results are given in section \ref{S:results} and conclusions are presented in section \ref{S:conclusions}.

\section{Methods} \label{S:methods}


We use a Cartesian coordinate system $(x,y,z)$ where the gravitational acceleration is constant and directed normal to the boundaries, $\boldsymbol{g}=-g \hz$, where $\hz$ is the vertical unit vector. 
The imposed horizontal magnetic field is defined by
\be
\mathbf{B}_0 = B_0 \hx ,
\ee
where $\hx$ is the unit vector pointing in the $x$ direction. Here $B_0$ is the constant dimensional amplitude of the imposed magnetic field; in dimensionless form the magnitude of the magnetic field is commonly specified by the Chandrasekhar number
\be
Q = \frac{B_0^2 H^2}{\rho \nu \mu_0 \eta} ,
\ee
where $\rho$ is the fluid density,  $\mu_0$ is the vacuum permeability and $\eta$ is the magnetic diffusivity. The fluid is assumed to be Oberbeck-Boussinesq. 

In the present study we assume that the magnetic Reynolds number $Rm = U H/\eta \rightarrow 0$. This asymptotic limit is referred to as the quasi-static approximation given that the induced magnetic field adjusts instantaneously to the velocity field when $Rm \ll 1$ is satisfied \citep[e.g.][]{bK08}. The quasi-static limit is equivalent to assuming that the induced magnetic field is asymptotically smaller than the imposed field; it is often an accurate approximation in liquid metal laboratory experiments in which the magnetic Prandtl number, $Pm = \nu / \eta = O(10^{-6})$, provided flow speeds are not too large \citep[e.g.][]{tV18,tZ20,yX22,tV21}.  

\textcolor{black}{We non-dimensionalize the governing equations using the fluid depth $H$, viscous diffusion time $H^2/\nu$, magnetic field $B_0$ and temperature $\Delta T$. The equations are then given by} 
\be
\partial_t \ub +\ub \cdot \nabla \ub  =   - \nabla p + \frac{Ra}{Pr} \vartheta \, \hz  + Q \, \dsx \bb +  \nabla^2 \ub,
\label{eq7}
\ee
\be
0=  \dsx \ub + \nabla^2\bb  ,
\label{eq8}
\ee
\be
\dst \vartheta + \ub \cdot \nabla \vartheta = \frac{1}{Pr} \lap \vartheta ,
\label{E:heat}
\ee
\be
\nabla \cdot \ub = 0, 
\ee
\be
\nabla \cdot \bb = 0,
\ee
where $\ub = (u, v, w)$ is the velocity vector, $p$ is the pressure, $\vartheta$ is the temperature and $\bb = (b_x, b_y, b_z)$ is the induced magnetic field vector. The mechanical boundary conditions are impenetrable
\be
w = 0 \quad \text{at} \quad z = 0, 1,
\ee
and stress-free,
\be
\dsz u = \dsz v = 0 \quad \text{at} \quad z = 0, 1.
\ee
The thermal boundary conditions are constant temperature,
\be
\vartheta = 1 \quad \text{at} \quad z = 0, \quad \text{and} \quad \vartheta = 0 \quad \text{at} \quad z = 1 .
\ee
The electromagnetic boundary conditions are electrically-insulating such that the magnetic field is matched to a potential field at the top and bottom boundaries \citep[e.g.][]{cJ00b}.

The equations are solved numerically with a pseudo-spectral code that uses Chebyshev polynomials in the vertical coordinate, Fourier series in the horizontal dimensions and a 3rd-order accurate implicit/explicit time-stepping scheme. The velocity and magnetic field vectors are represented in terms of poloidal and toroidal scalars such that the solenoidal conditions are satisfied exactly \citep{pM16}. 

Only computational domains of square cross-section are used in the present work. The horizontal periodicity length, $L$, is scaled by the non-dimensional critical horizontal wavelength, $\lambda_c$, such that the aspect ratio of the domain is given by
\be
\Gamma \equiv \frac{L}{H} = n \lambda_c ,
\ee
where $n$ is the number of horizontal critical wavelengths present within the domain. For stress-free mechanical boundary conditions we have $\Gamma \approx 2.83 n$. We found that $n=6$ is sufficient to obtain convergence of global statistics, while still remaining computationally feasible for simulating a wide range of parameter values. Thus, the aspect ratio used here is fixed at $\Gamma \approx 17$ for all simulations.


Some previous studies of MC use the horizontal length scale $L$ of the domain in the definition of the Chandrasekhar number \citep[e.g.][]{tV21}.  If we denote $Q_L$ as the Chandrasekhar number based on $L$ then  we have
\be
Q_L = \frac{B_0^2 L^2}{\rho \nu \mu_0 \eta} = \Gamma^2 Q .
\ee
Thus, for the aspect ratio used here we have $Q_L \approx 288 Q$. This disparity between the two different definitions should be noted when comparing with previous work.  $Q_L$ becomes particularly relevant when vertical sidewalls contain the fluid, since Hartmann layers form and can have significant effects on the resulting convection \citep[e.g.][]{uB02}.



A parameter survey was carried out with $Q=[10^2, 10^3, 10^4, 10^5, 10^6]$ and Rayleigh numbers up to $Ra = 3 \times 10^7$. For computational simplicity we use a thermal Prandtl number of $Pr=1$. Where possible we compare with the hydrodynamic ($Q=0$) data of Ref.~\citep{mY19}. Simulations with $Q = [10^2, 10^4, 10^6]$ were conducted to systematically explore the parameter space of the MC system. A subset of simulations with $Q = [10^3, 10^5]$ were conducted primarily to investigate the transition from 2D to 3D regimes, and therefore the range of investigated Rayleigh numbers for these cases was comparatively smaller. In terms of simulation progression, $Ra$ is increased incrementally for each fixed value of $Q$, and all simulations are run until a statistically stationary state is reached. The details of each simulation are listed in Table \ref{T:data}. 

%

Various output quantities are used to analyze the results of the simulations. All time and volume averaged quantities are denoted with angled brackets and quantities averaged over horizontal planes are denoted with an overline. The Nusselt number, which measures the non-dimensional heat transport, is defined as
\be
Nu = 1 + Pr \langle w \vartheta' \rangle ,
\ee
where $\vartheta' = \vartheta - \overline{\vartheta}$ is the fluctuating temperature field. 
With our non-dimensionalization the Reynolds number is defined as 
\be
Re = \sqrt{\langle \ub \cdot \ub \rangle}.
\ee
We also find it useful to define the Reynolds numbers based on the transverse ($y$) and vertical ($z$) components of the velocity, i.e.~
\be
Re_y = \sqrt{ \langle v'^2 \rangle }, \qquad Re_z = \sqrt{ \langle w^2 \rangle },
\ee
where $v' = v - \overline{v}$ is the transverse component of the fluctuating velocity.



The exact relationship between the heat transport and the two sources of dissipation (viscous and ohmic) can be derived from the governing equations to give \citep[e.g.][]{sC61},
\be
\frac{Ra}{Pr^2} \lb Nu - 1 \rb  = \varepsilon_u + \varepsilon_b ,
\label{E:Nu_diss}
\ee
where the viscous and ohmic dissipation \textcolor{black}{rates} are defined by, respectively,
\be
\varepsilon_u = \langle |\nabla \times \ub|^2 \rangle, \quad \text{and} \quad \varepsilon_b = Q \langle |\nabla \times \bb|^2 \rangle .
\ee
With these definitions we also find it useful to compute the fraction of ohmic dissipation,
\be
f_{\Omega} = \frac{\varepsilon_b}{\varepsilon_u + \varepsilon_b}.
\ee 


\section{Results}
\label{S:results}

\subsection{Flow structure}

\begin{figure}
 \begin{center}
      \subfloat[][]{\includegraphics[width=0.48\textwidth]{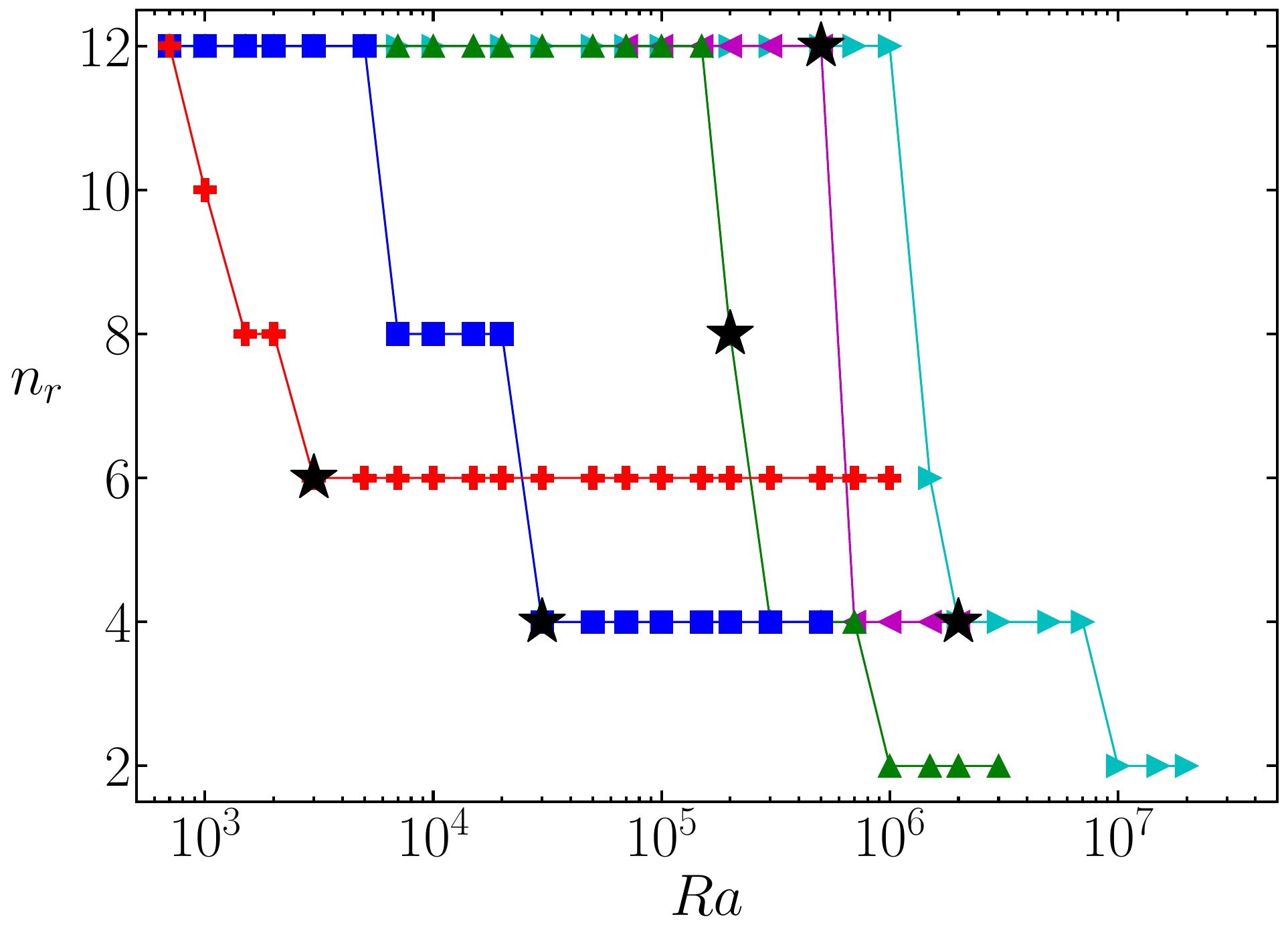}} \quad
      \subfloat[][]{\includegraphics[width=0.48\textwidth]{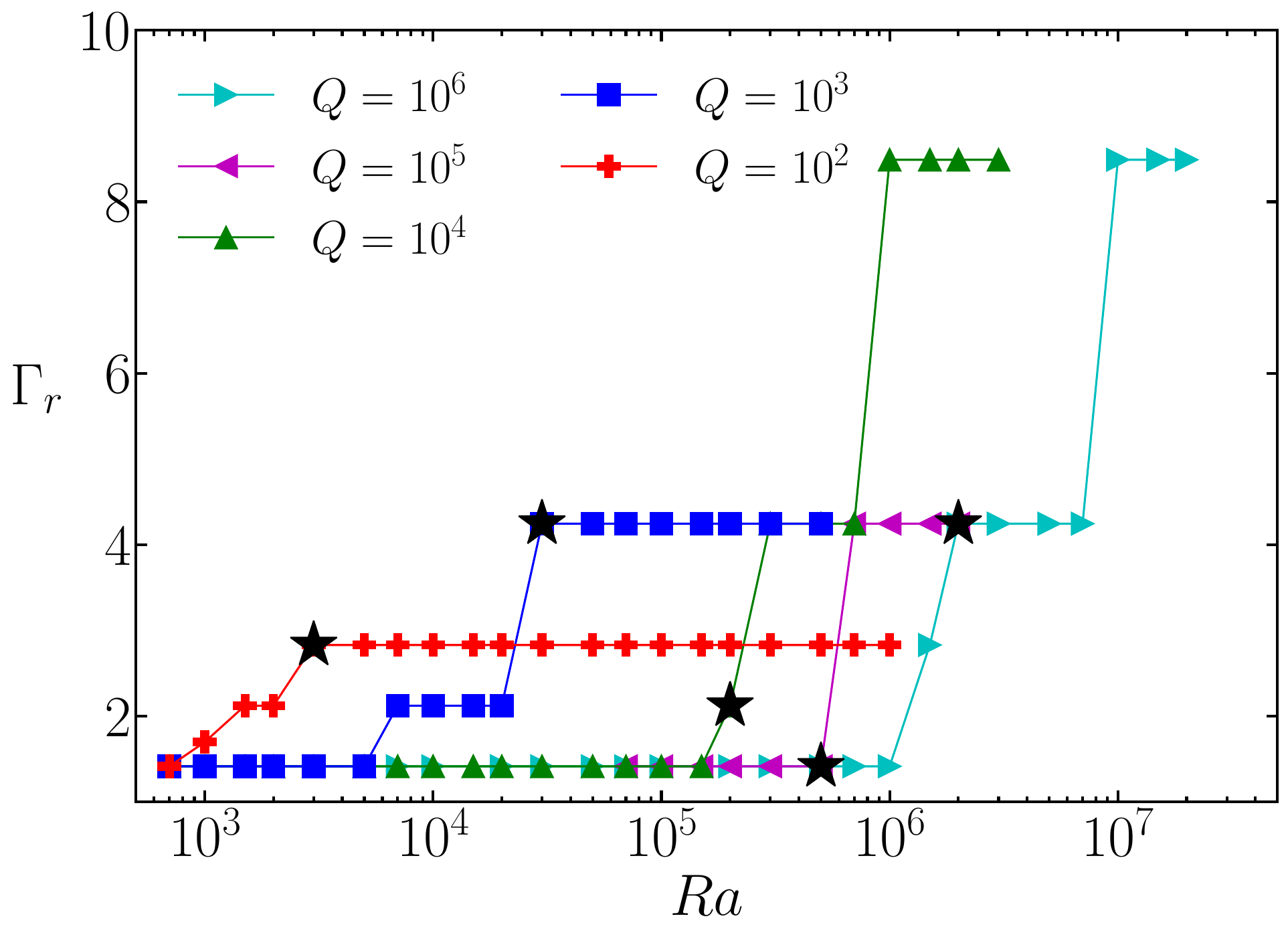}}
      \caption{Convective structure versus Rayleigh number ($Ra$): (a) number of convection rolls, $n_r$; and (b) convection roll aspect ratio, $\Gamma_{r}$. Black stars denote the values of $Ra$ for which three-dimensional motion is first observed for each value of $Q$.}
      \label{F:cells}
\end{center}
\end{figure}

As an initial characterization of the different states of flow observed in this system, we show the number of convection rolls ($n_r$) and the corresponding roll aspect ratio ($\Gamma_r$) for all of the simulations in Figs.~\ref{F:cells}(a) and (b), respectively.
These two quantities are related simply via
\be
\Gamma_{r} = \frac{\Gamma}{n_r}.
\ee
For unsteady cases, the values plotted correspond to time-averaged values. However, even for turbulent cases (e.g.~$Q=10^2$, $Ra=10^6$) there is significant anisotropy present in the flow such that a dominant roll structure persists. At the onset of convection our aspect ratio allows for $n_r = 12$ rolls, corresponding to $\Gamma_r \approx 1.4$. All values of $Q$ show a trend in which $n_r$ ($\Gamma_r$) decreases (increases) with increasing $Ra$. For $Q=10^4$ and $Q=10^6$ the flow eventually transitions to a state consisting of two convection rolls with large aspect ratio, $\Gamma_r \approx 8.5$. In general we find that fewer numbers of rolls, and therefore larger aspect ratio rolls become preferred as both $Ra$ and $Q$ are increased.

\begin{figure}
 \begin{center}
     \subfloat[]{\label{fig:left}%
      \begin{tabular}[b]{c}
        \includegraphics[width=0.22\textwidth]{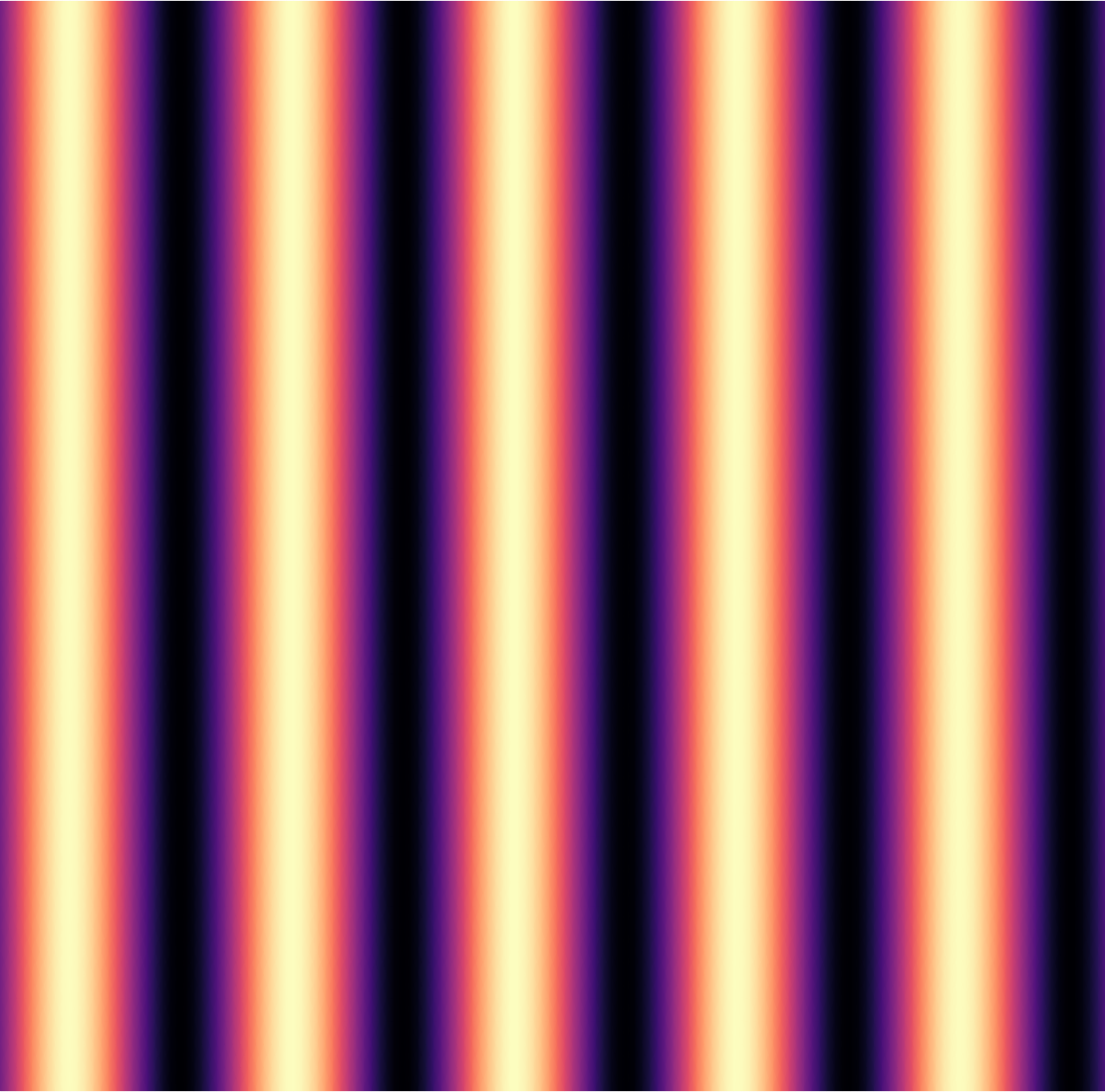}\\[1mm]
        \includegraphics[width=0.22\textwidth]{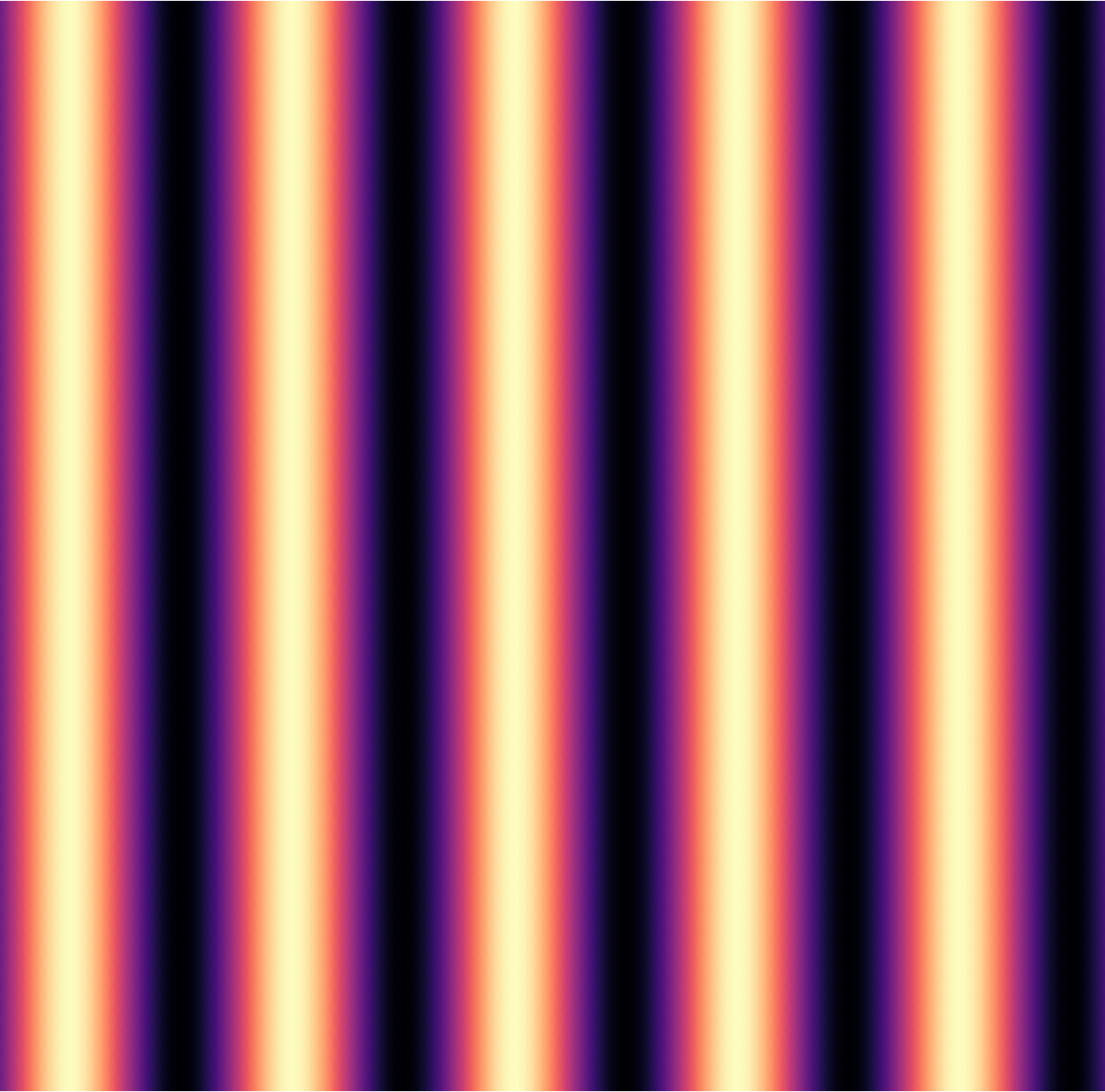}
      \end{tabular}}
     \subfloat[]{\label{fig:left}%
      \begin{tabular}[b]{c}
        \includegraphics[width=0.22\textwidth]{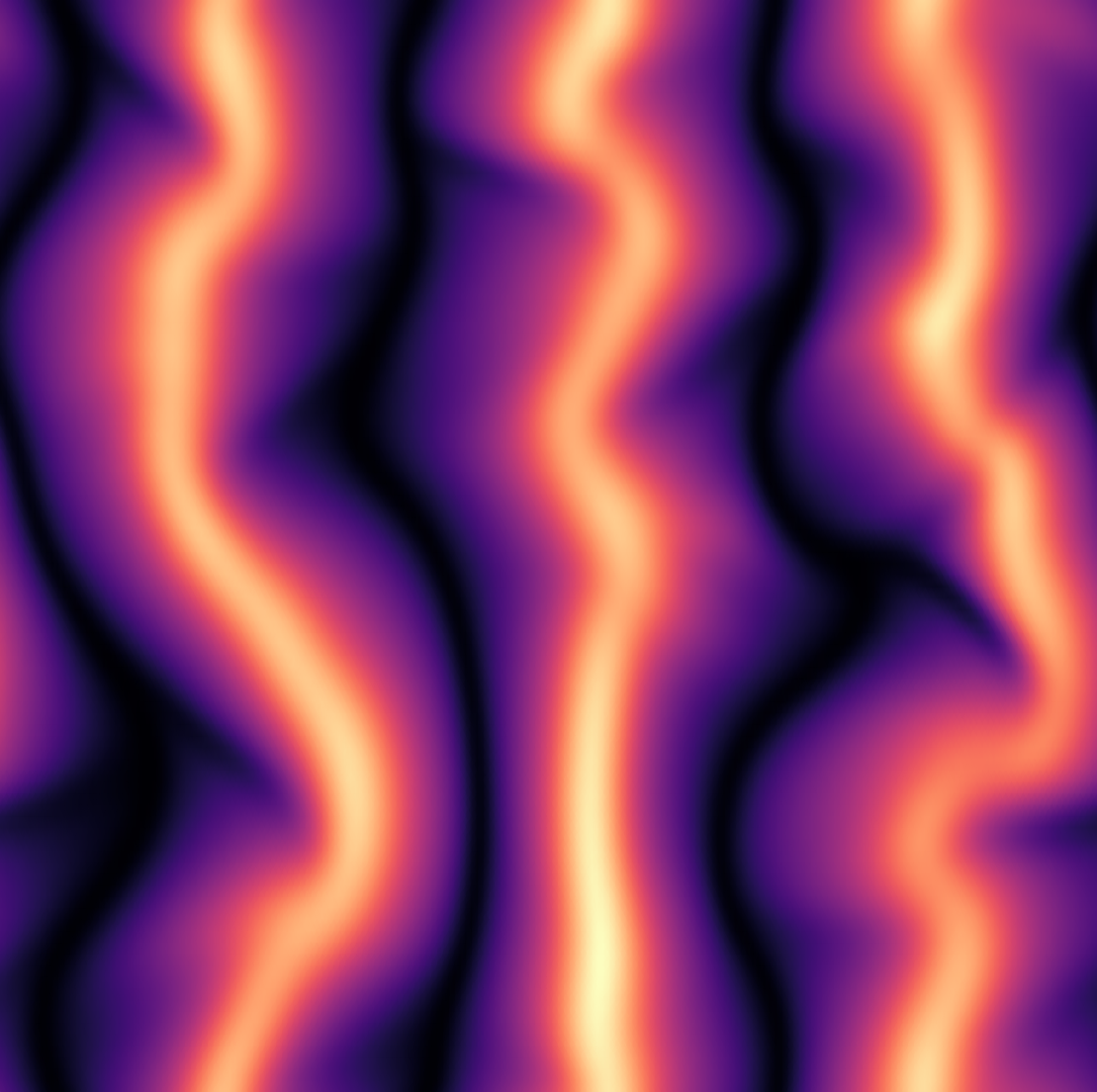}\\[1mm]
        \includegraphics[width=0.22\textwidth]{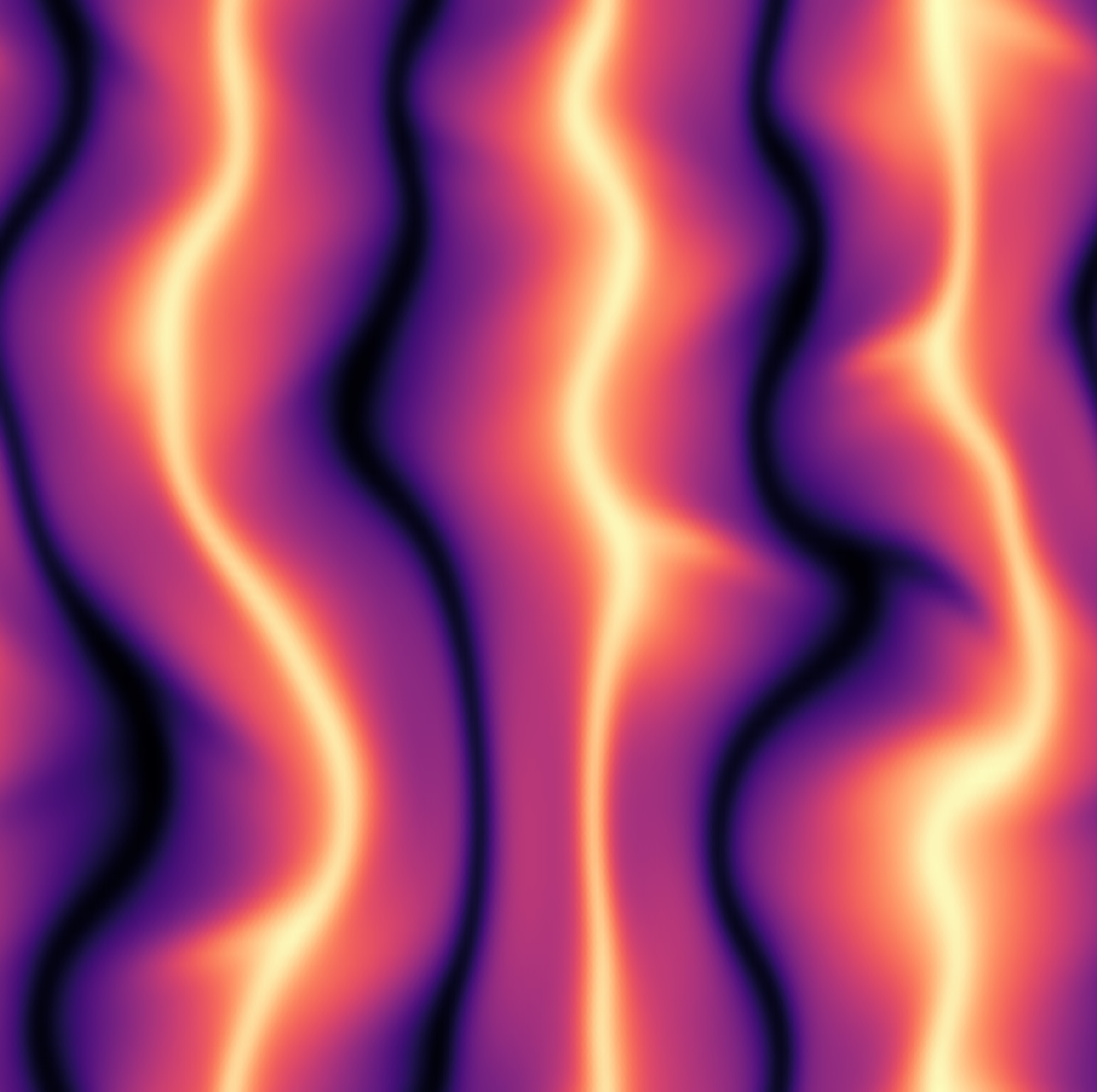}
      \end{tabular}}         
     \subfloat[]{\label{fig:left}%
      \begin{tabular}[b]{c}
        \includegraphics[width=0.22\textwidth]{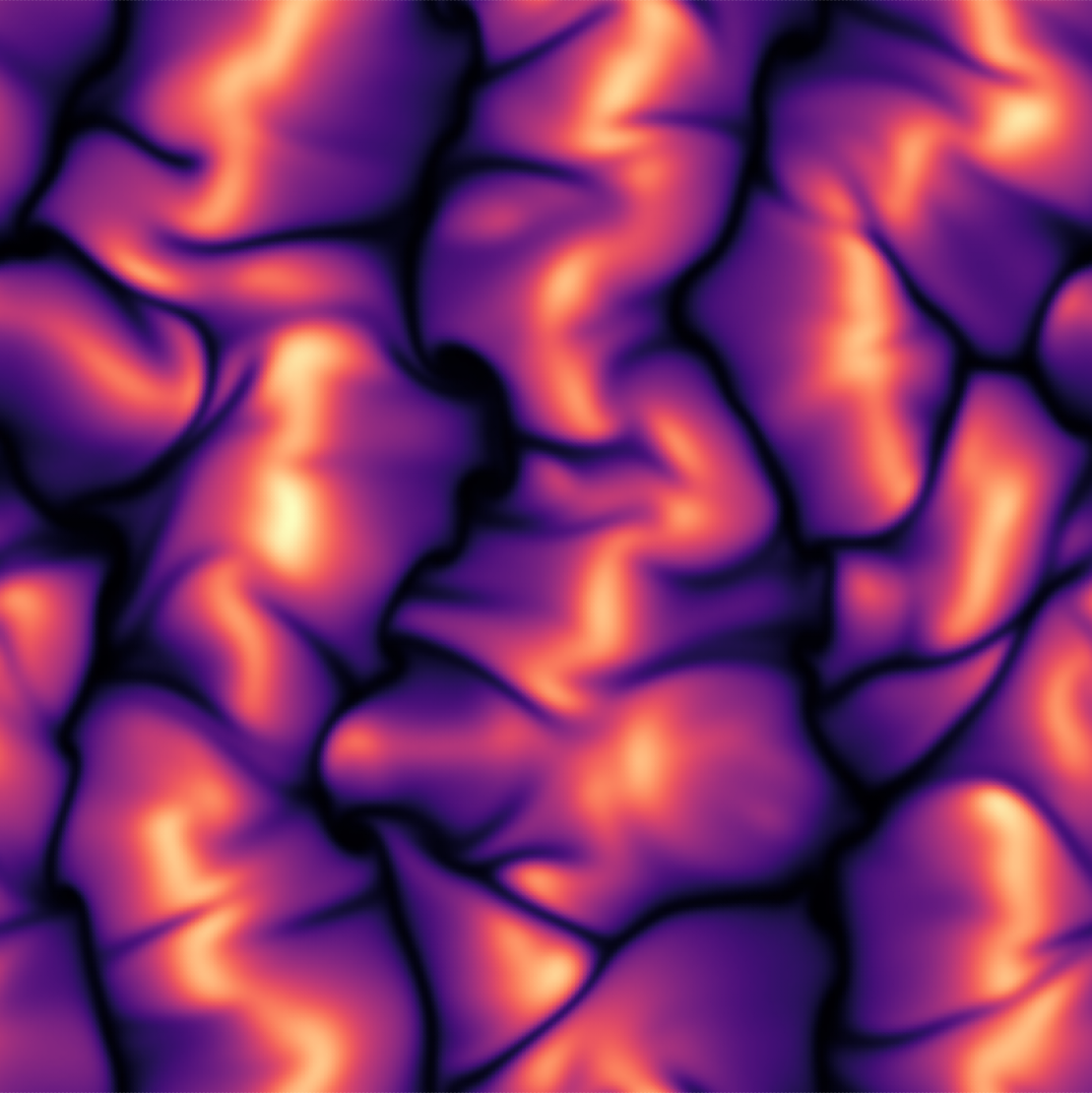}\\[1mm]
        \includegraphics[width=0.22\textwidth]{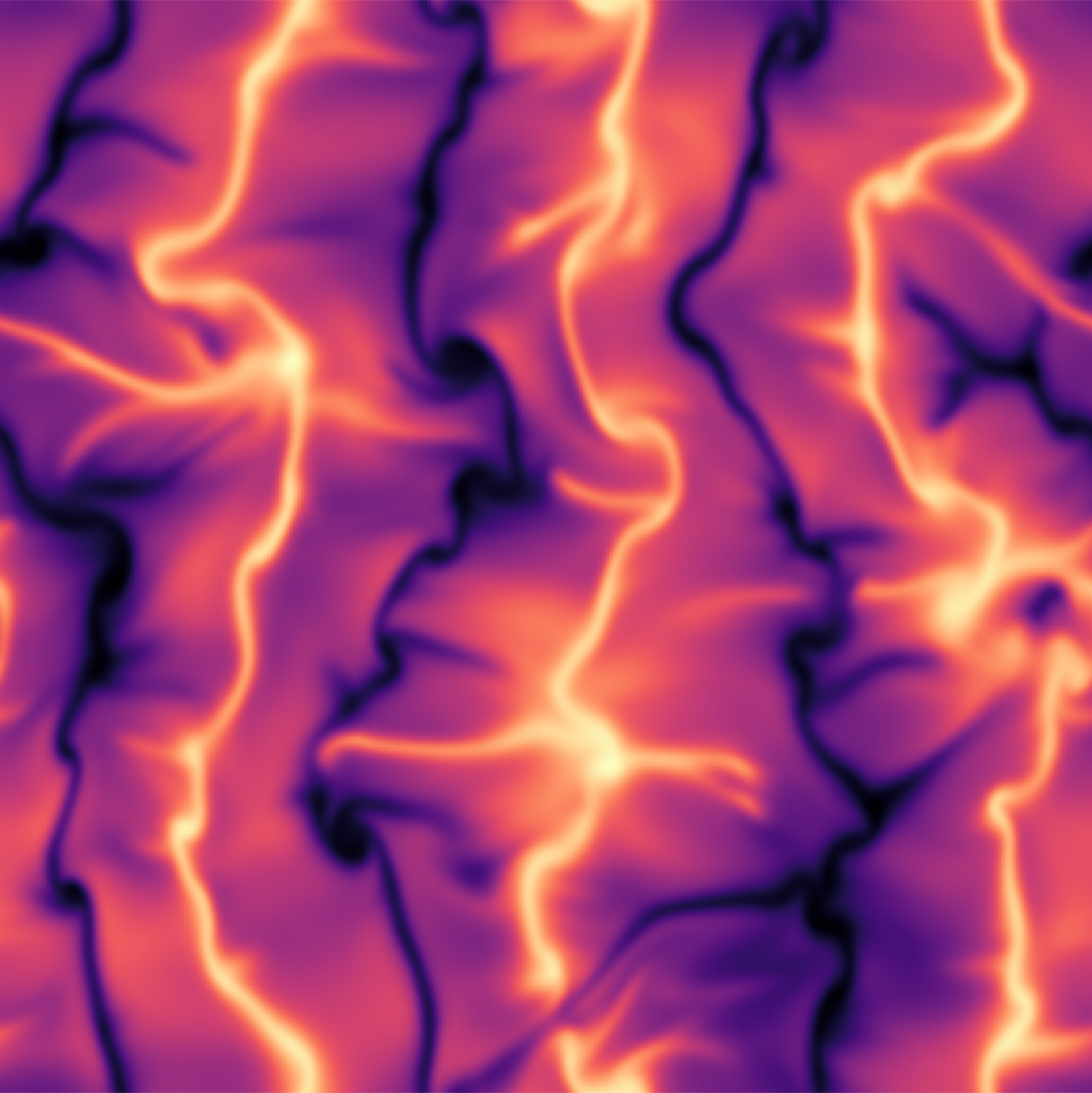}
      \end{tabular}}      
     \subfloat[]{\label{fig:left}%
      \begin{tabular}[b]{c}
        \includegraphics[width=0.22\textwidth]{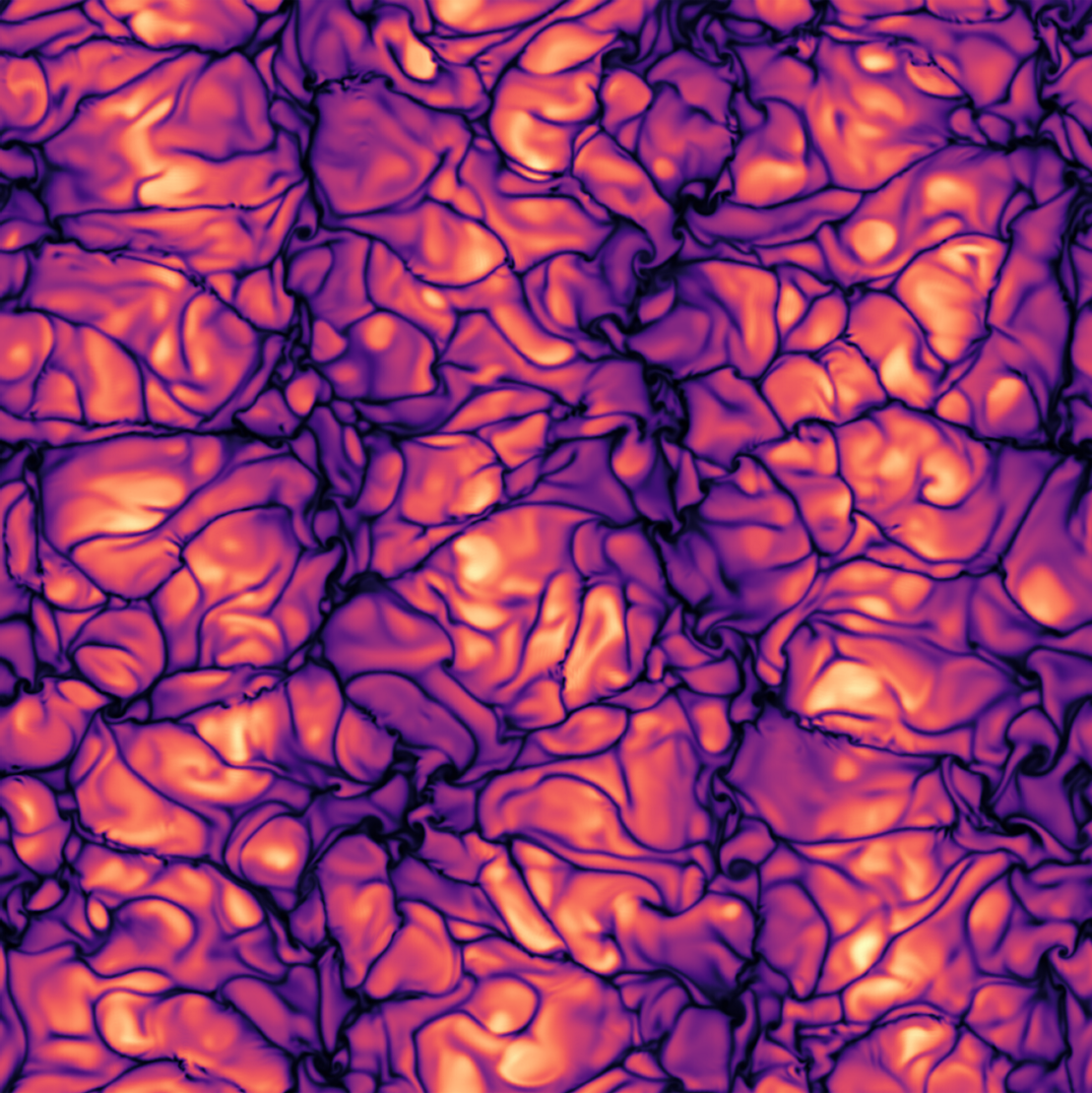}\\[1mm]
        \includegraphics[width=0.22\textwidth]{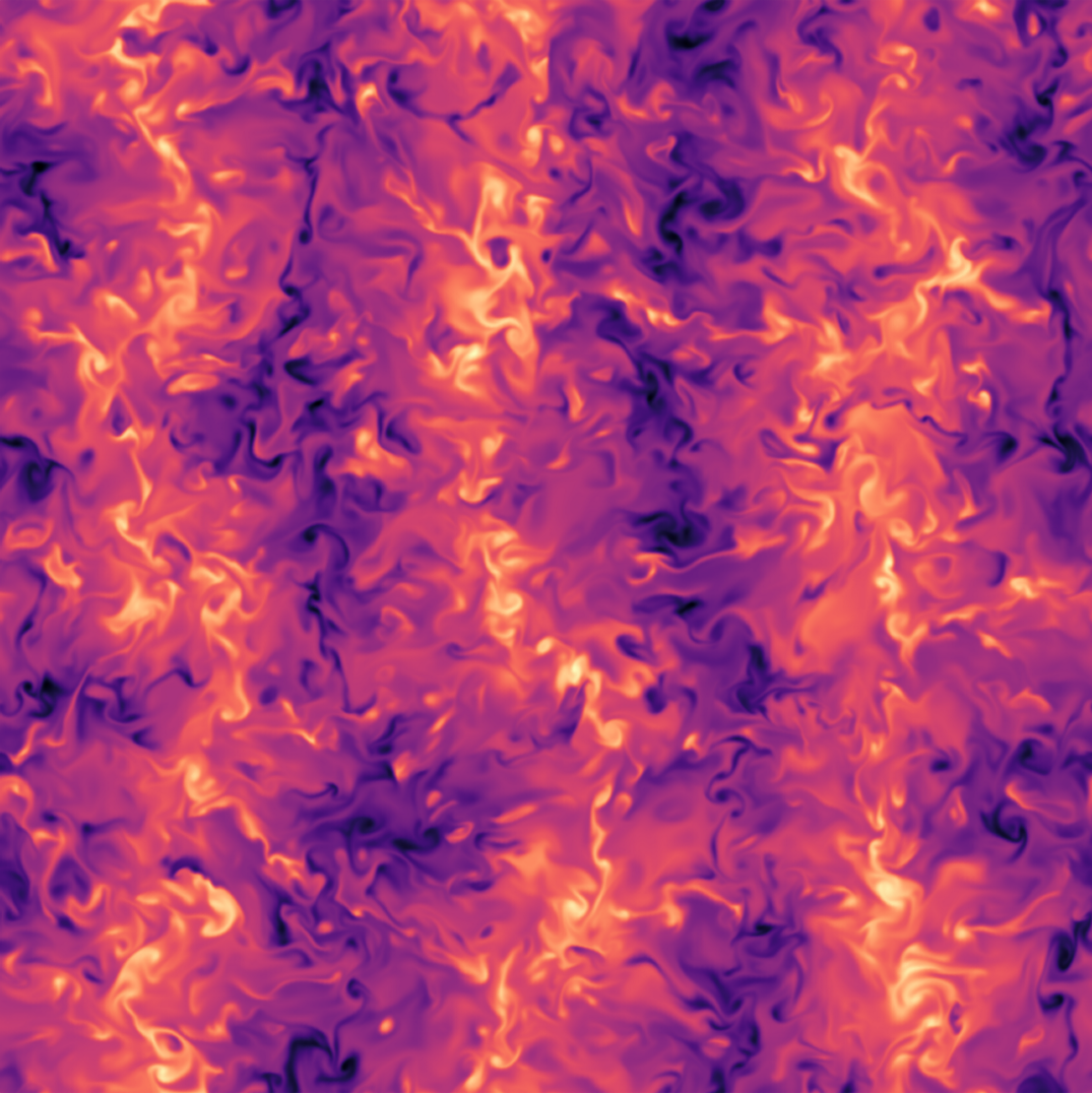} 
      \end{tabular}}         \\
      
     \subfloat[]{\label{fig:left}%
      \begin{tabular}[b]{c}
        \includegraphics[width=0.225\textwidth]{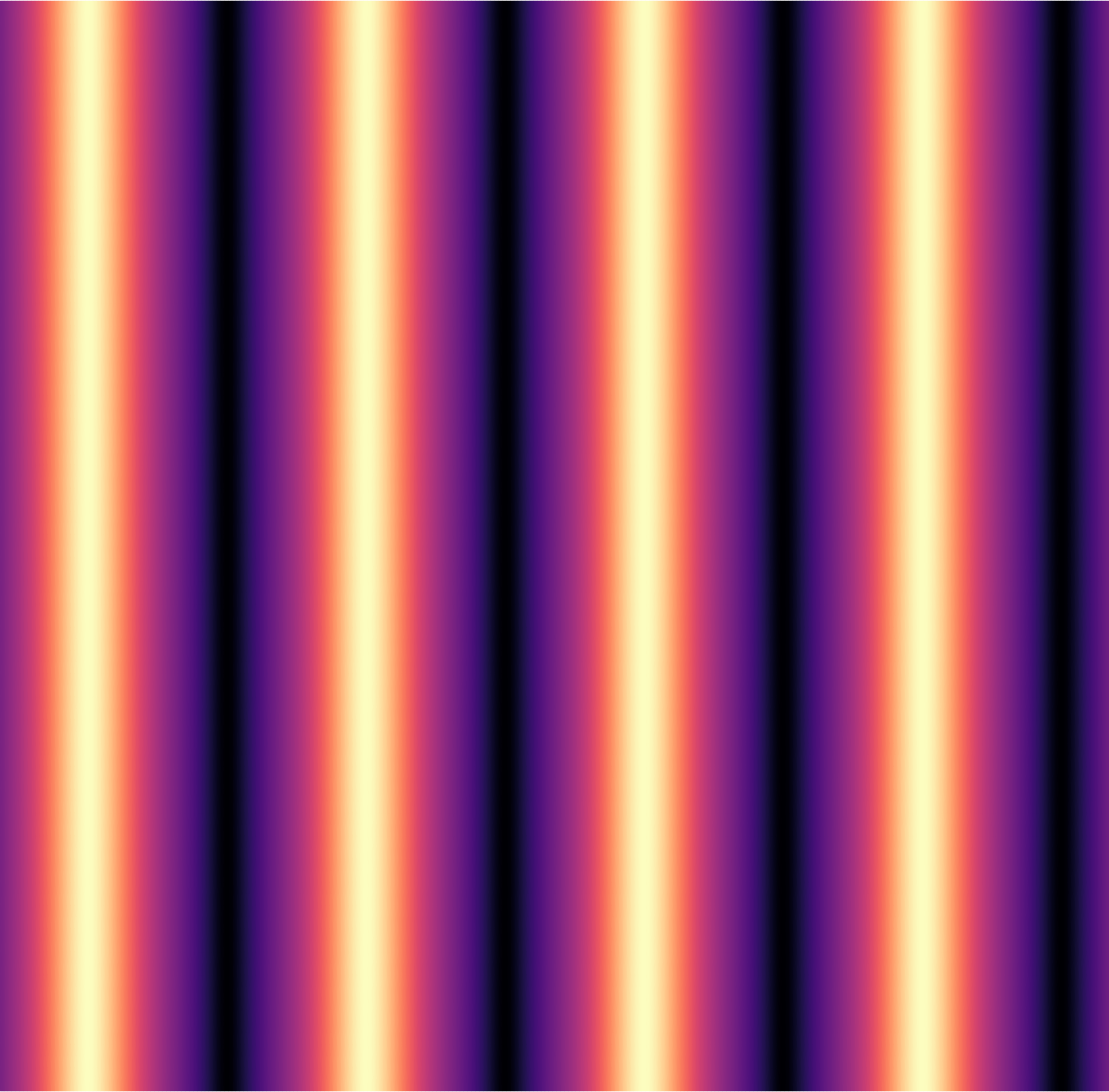}\\[1mm]
        \includegraphics[width=0.225\textwidth]{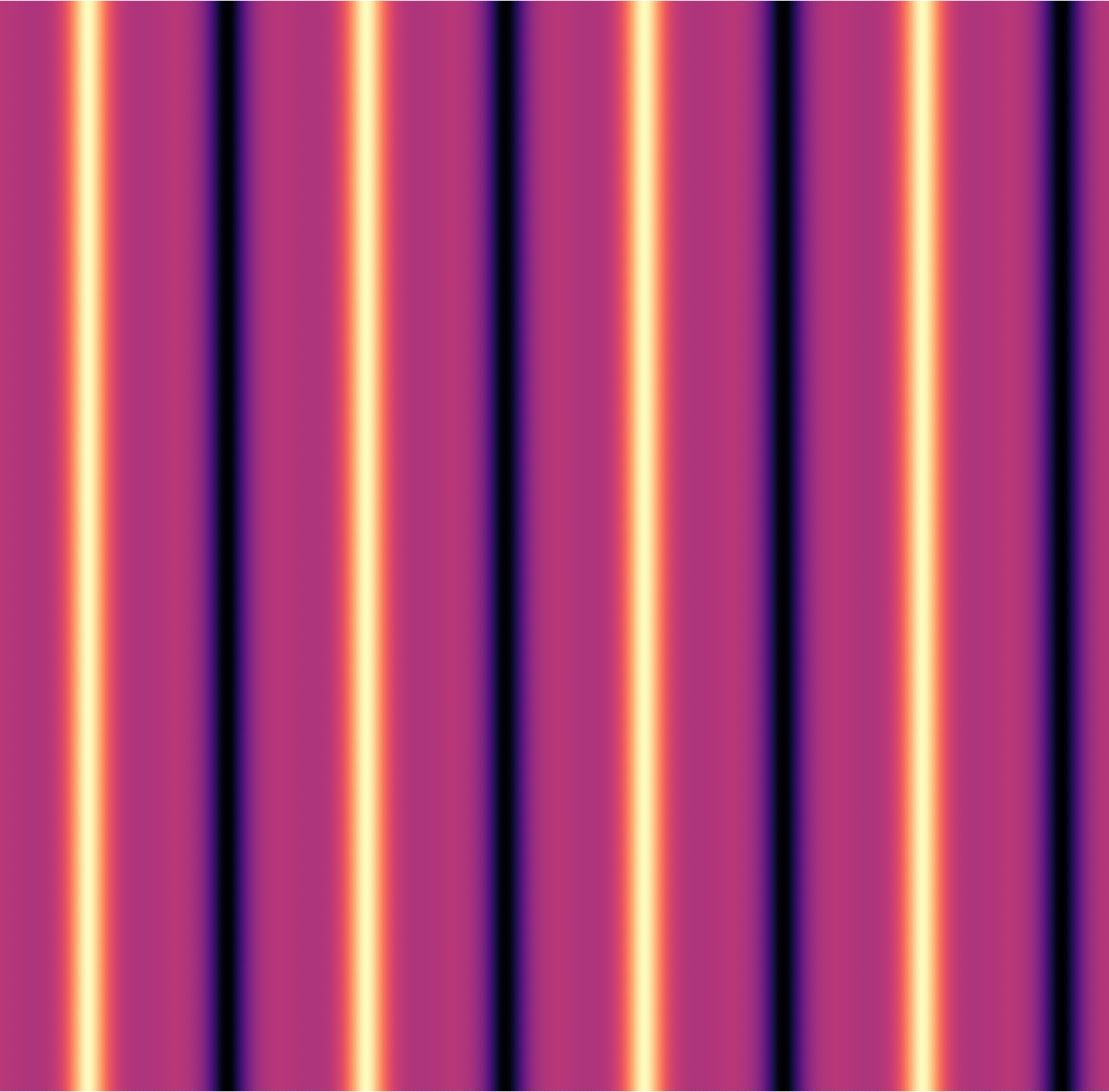}
      \end{tabular}}
     \subfloat[]{\label{fig:left}%
      \begin{tabular}[b]{c}
        \includegraphics[width=0.225\textwidth]{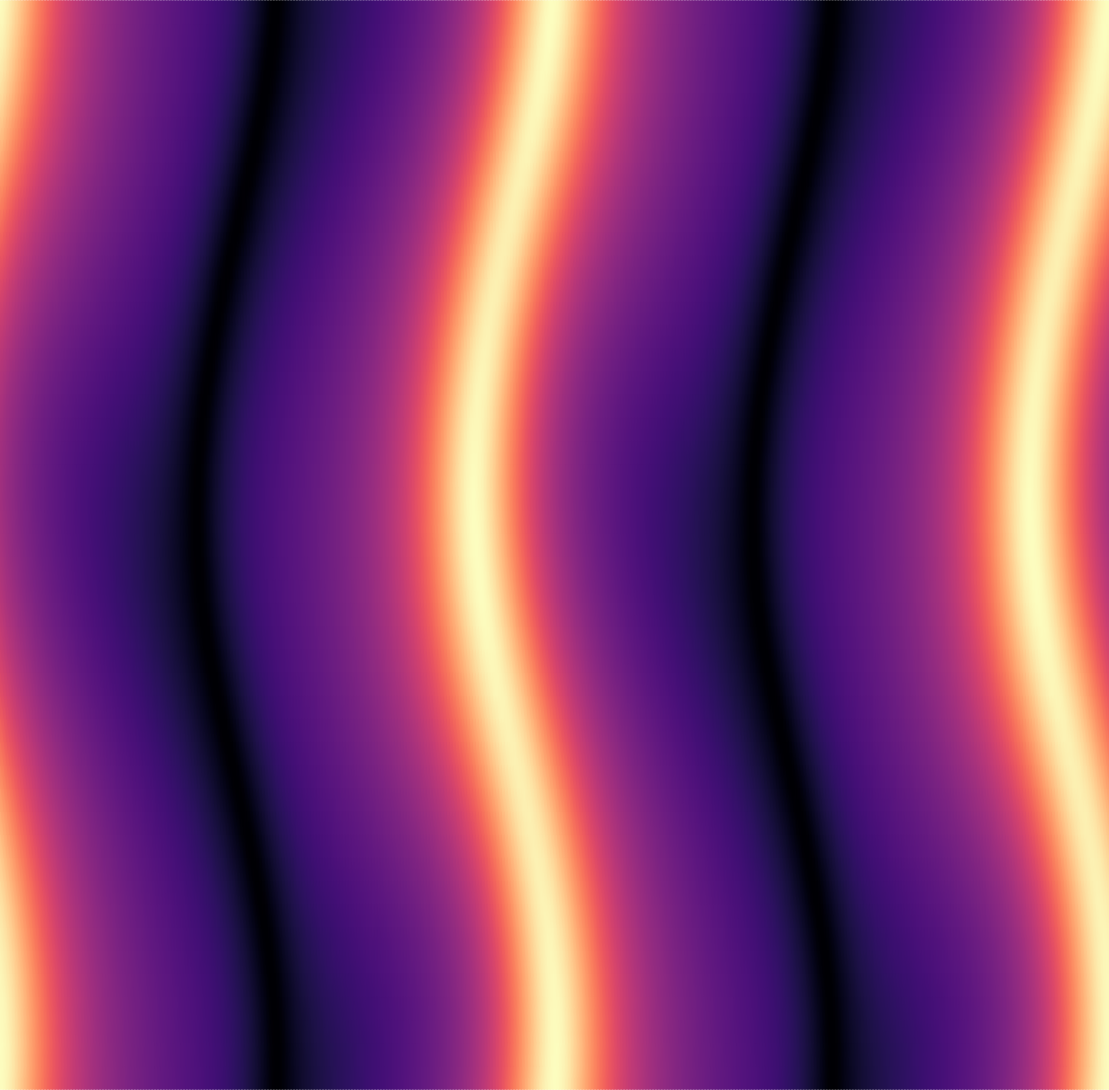}\\[1mm]
        \includegraphics[width=0.225\textwidth]{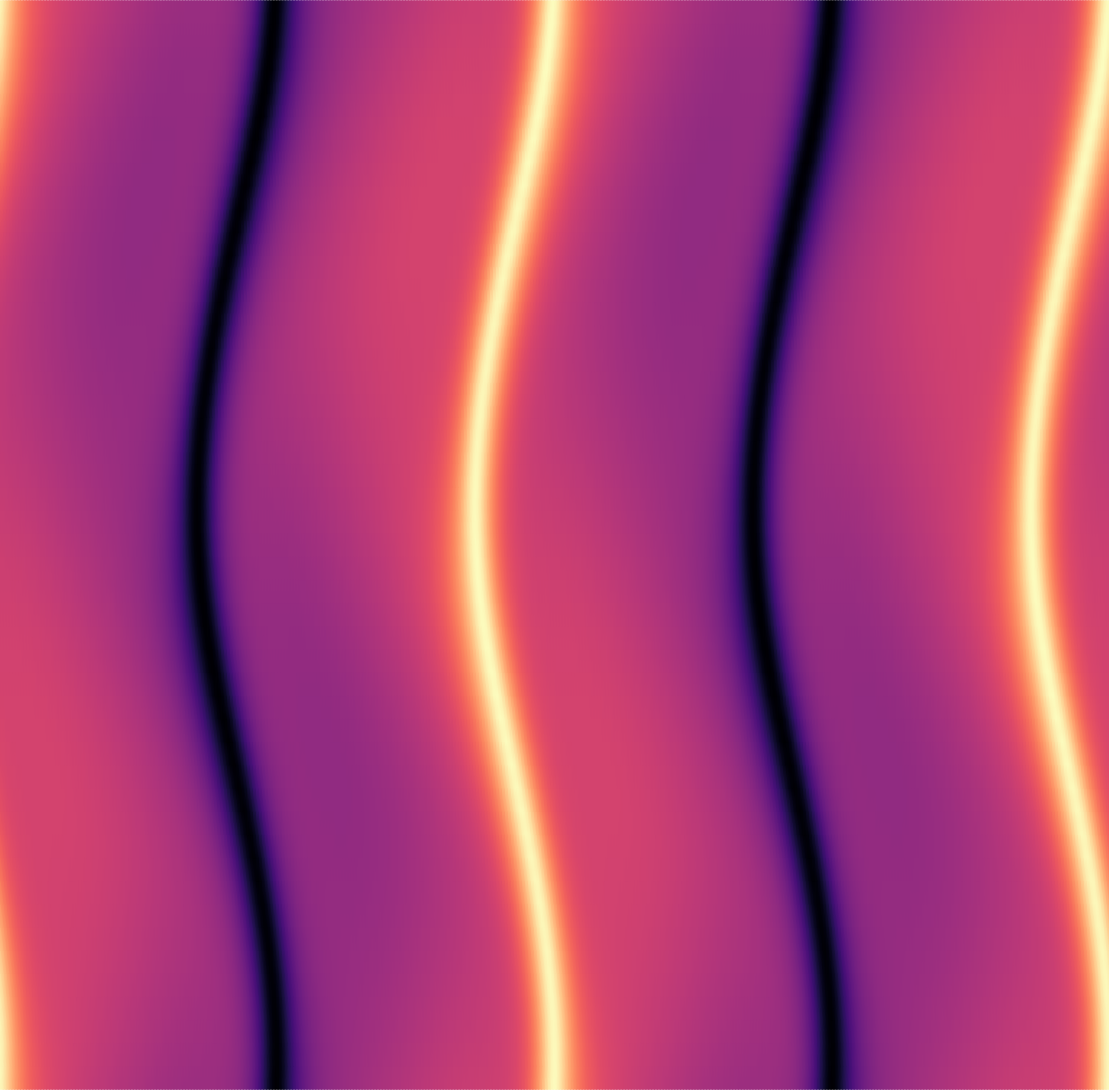} 
      \end{tabular}}      
     \subfloat[]{\label{fig:left}%
      \begin{tabular}[b]{c}
        \includegraphics[width=0.22\textwidth]{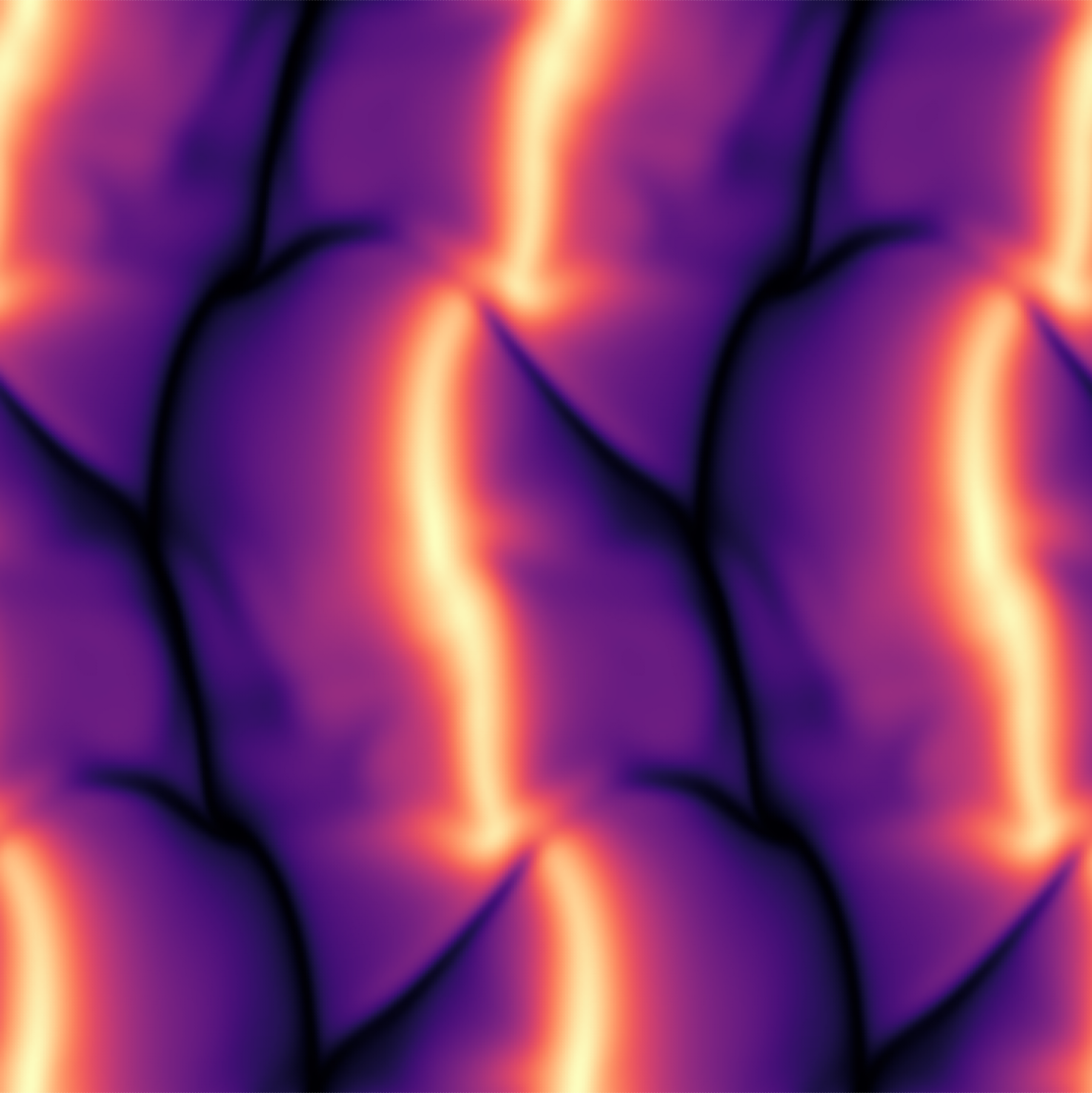}\\[1mm]
        \includegraphics[width=0.22\textwidth]{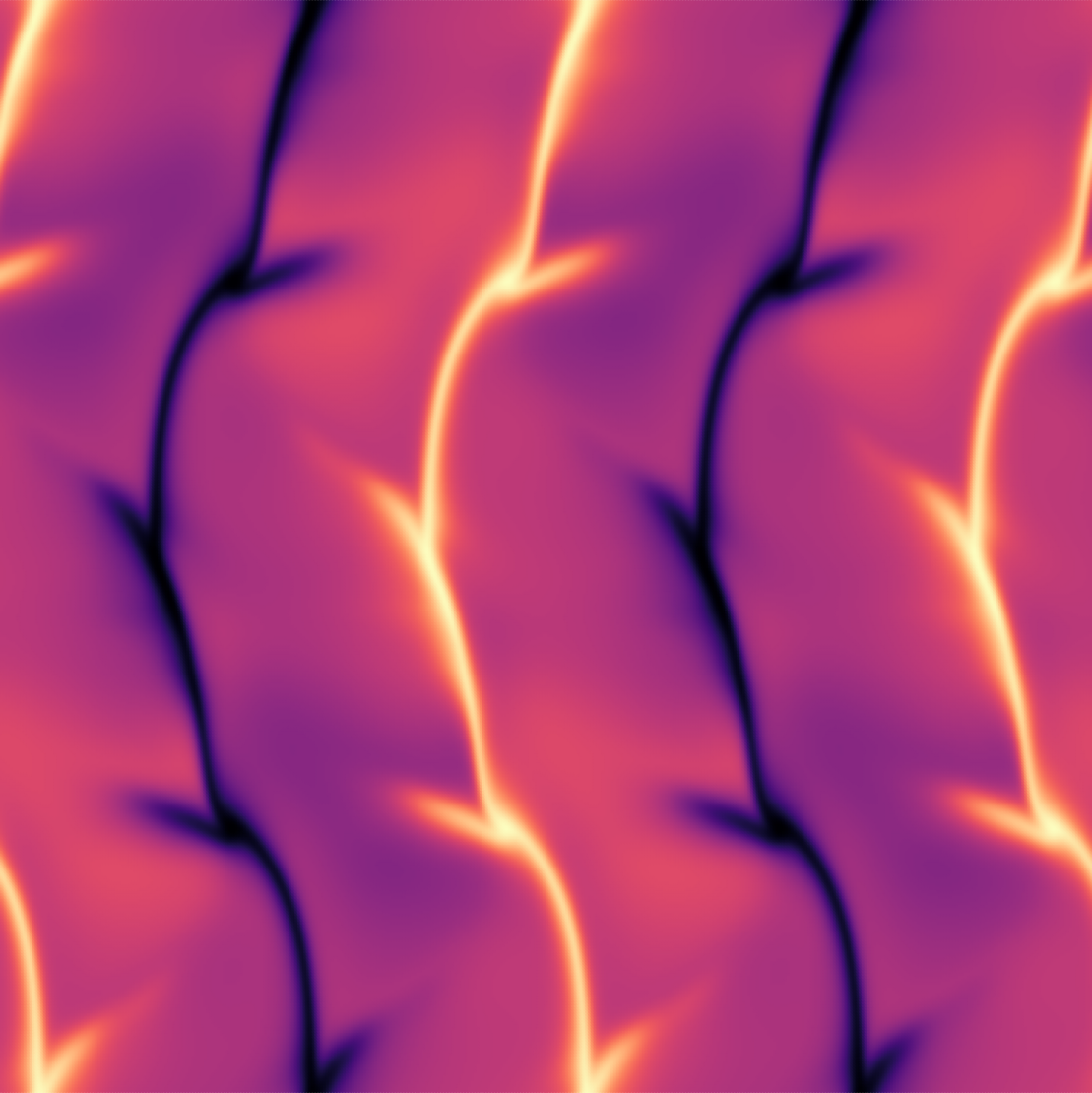}
      \end{tabular}}       
      \subfloat[]{\label{fig:left}%
      \begin{tabular}[b]{c}
        \includegraphics[width=0.22\textwidth]{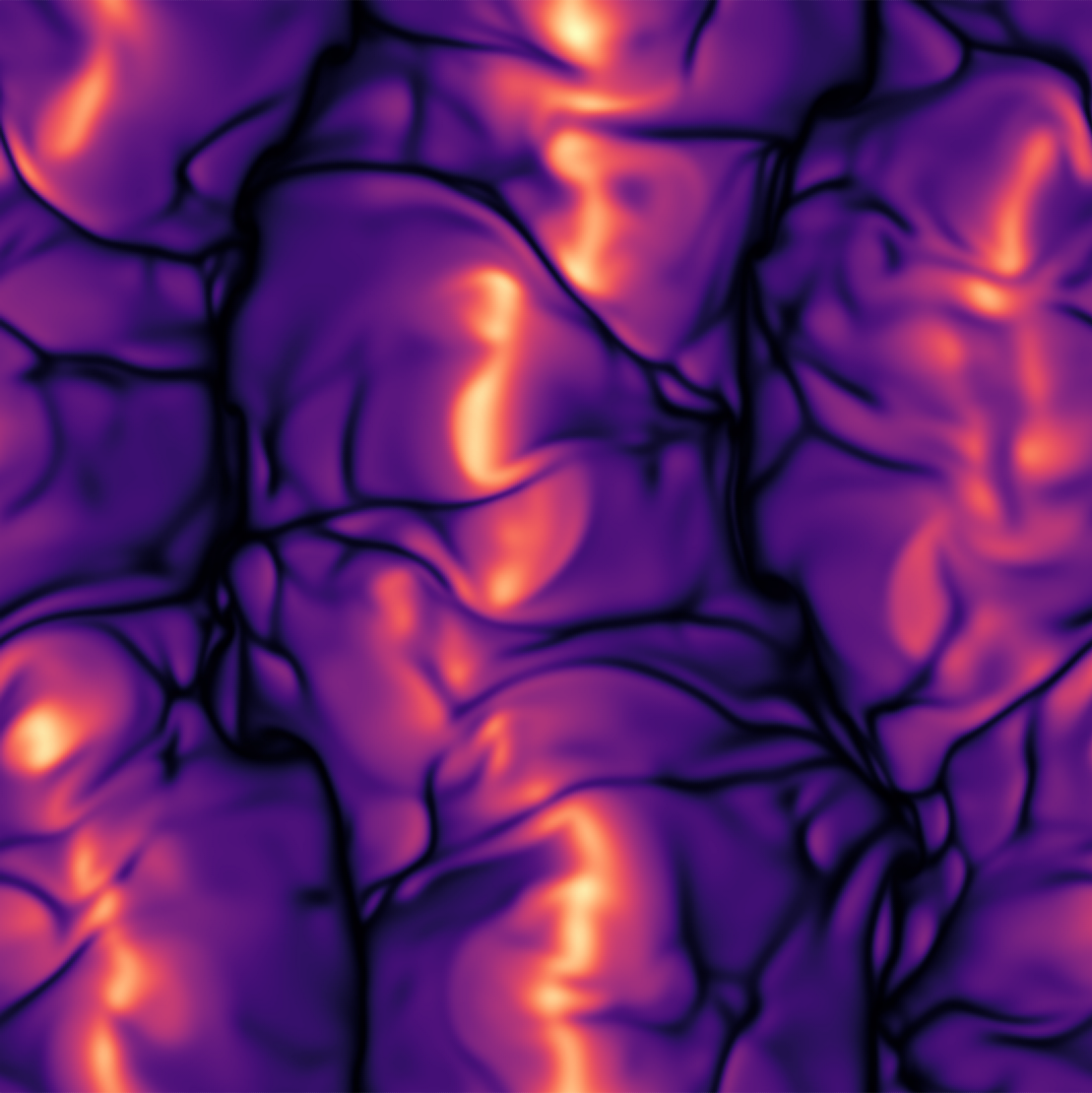}\\[1mm]
        \includegraphics[width=0.22\textwidth]{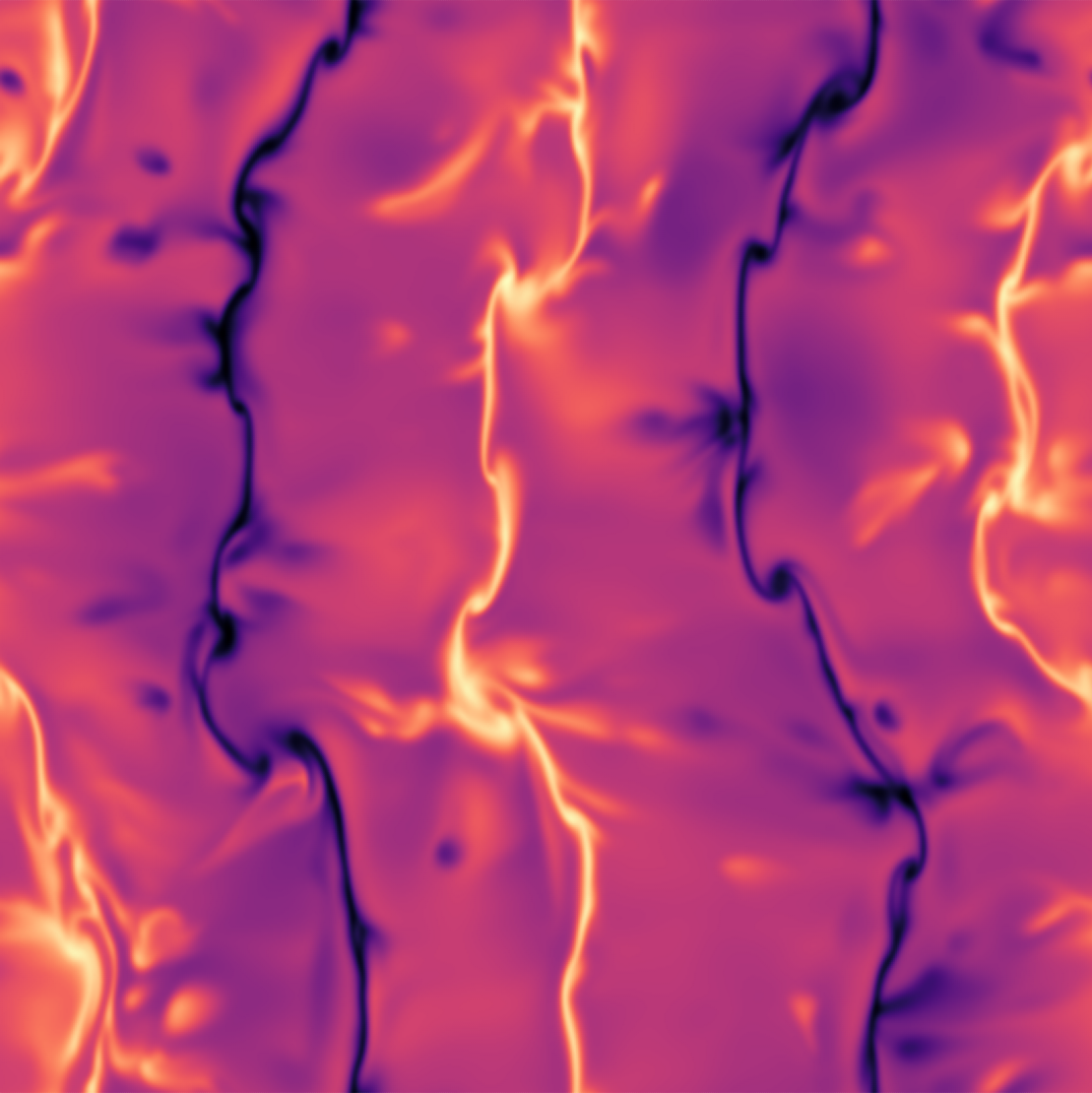}
      \end{tabular}}         
      \caption{Horizontal slices of the fluctuating temperature at the thermal boundary layer depth (top rows) and the midplane (bottom rows) for (a)-(d) $Q=10^2$ and (e)-(h) $Q=10^3$. The Rayleigh number increases from left to right. The imposed magnetic field points up the page; gravity points into the page. Perspective views of volumetric renderings of the fluctuating temperature illustrating the four primary flow regimes identified: (a) the weakly magnetic regime ($Q = 10^2, \, Ra = 2 \times 10^4$)}
      \label{F:flow}
\end{center}
\end{figure}

Visualizations from a sample of the different flow regimes are given in Fig.~\ref{F:flow}, where snapshots of horizontal slices of the temperature perturbation are shown through both the thermal boundary layer (top row in (a)-(d) and top row in (e)-(h)) and the midplane  (bottom row in (a)-(d) and bottom row in (e)-(h)) for $Q=10^2$ ((a)-(d)) and $Q=10^3$ ((e)-(h)). The Rayleigh number increases from left to right within each row. Recall that we use the term `two-dimensional' (2D) to refer to flows that are independent of $x$, i.e.~invariant in the direction of the imposed magnetic field.
For $Q=10^2$, in panel (a) we show a steady 2D state with $n_r=10$ at $Ra=10^3$. At $Ra=7\times10^3$, the flow shown in panel (b) consists of six rolls that are unsteady and 3D. This same number of rolls persists at all of the investigated values of $Ra$ for $Q=10^2$, as shown in panels (c) and (d) (and Fig.~1(a)), which show the flow for $Ra=3\times10^4$ and $Ra=10^6$, respectively. Qualitatively similar transitions in the flow regimes occur for $Q=10^3$, where we show a steady 2D, $n_r=8$ regime at $Ra=10^4$ in panel (e); a steady, wavy (3D) regime in panel (f) at $Ra=3\times10^4$; the first unsteady case for $Ra=10^5$ is given in (g); and a case that shows significant chaotic temporal fluctuations is shown in (h) for $Ra=3\times10^5$.

\begin{figure}
 \begin{center}
     \subfloat[]{\label{fig:left}%
      \begin{tabular}[b]{c}
        \includegraphics[width=0.22\textwidth]{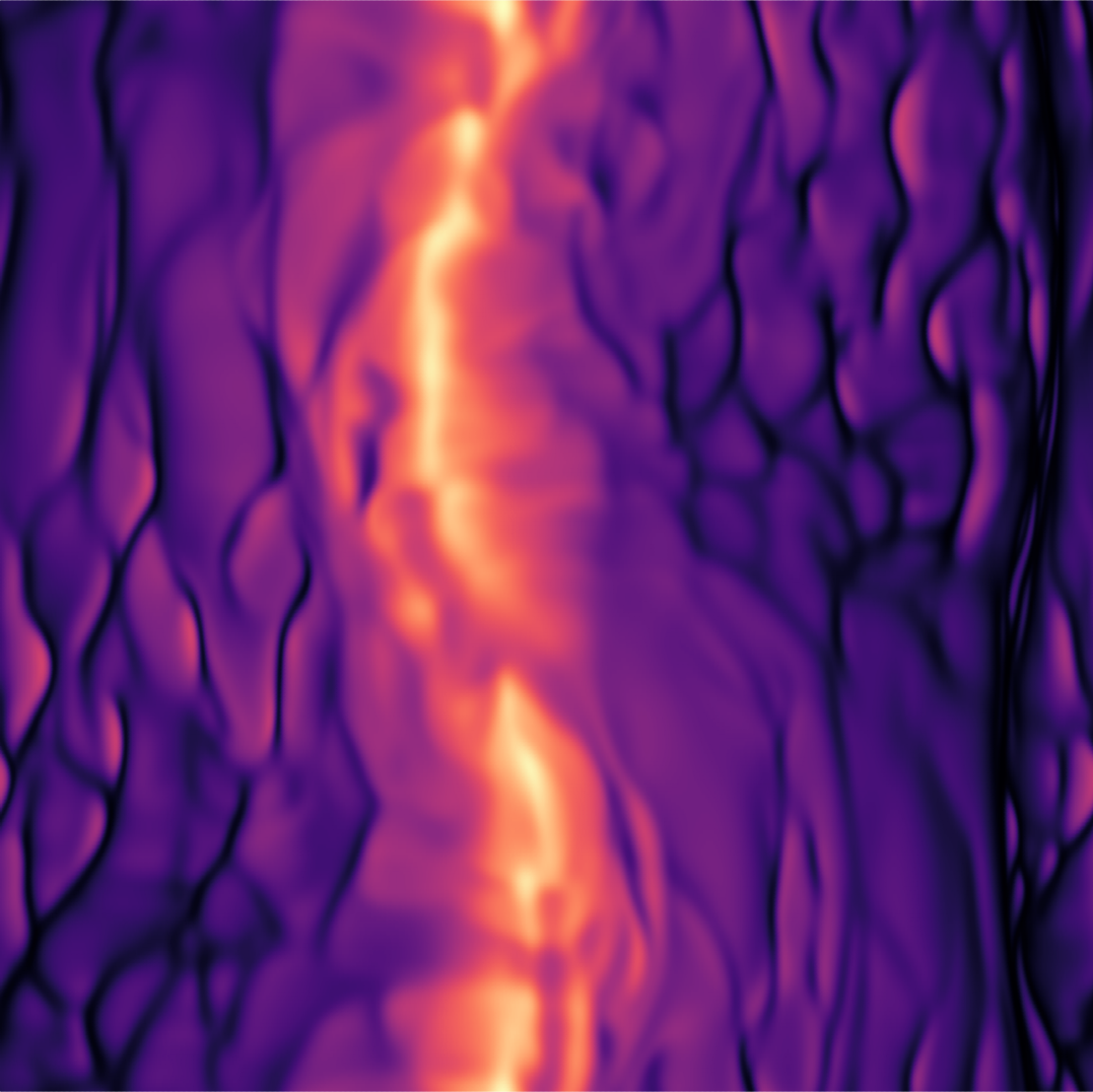}\\[1mm]
        \includegraphics[width=0.22\textwidth]{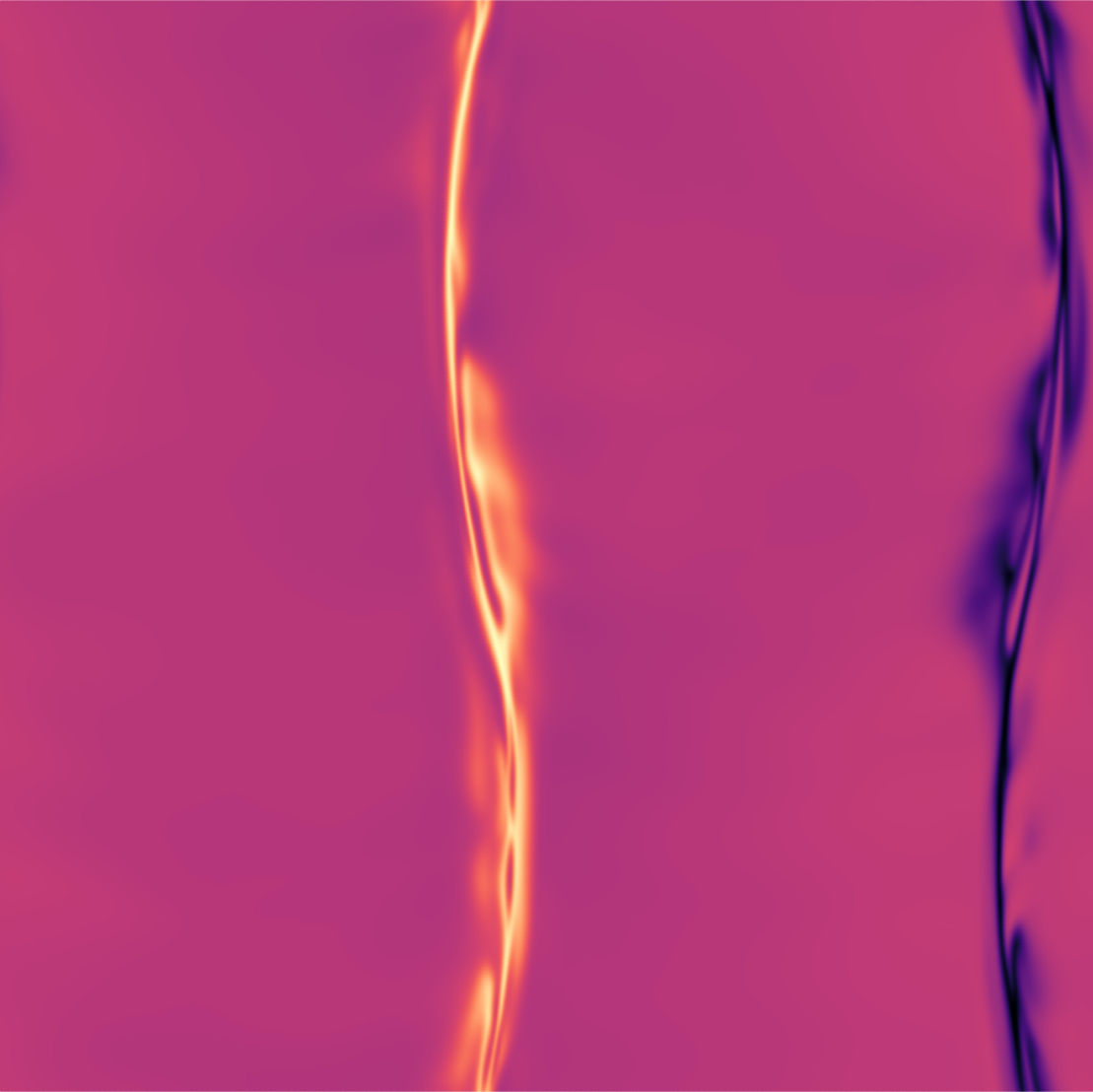}\\[1mm]
      \end{tabular}}      
      \qquad 
      \subfloat[]{\label{fig:left}%
      \begin{tabular}[b]{c}
        \includegraphics[width=0.22\textwidth]{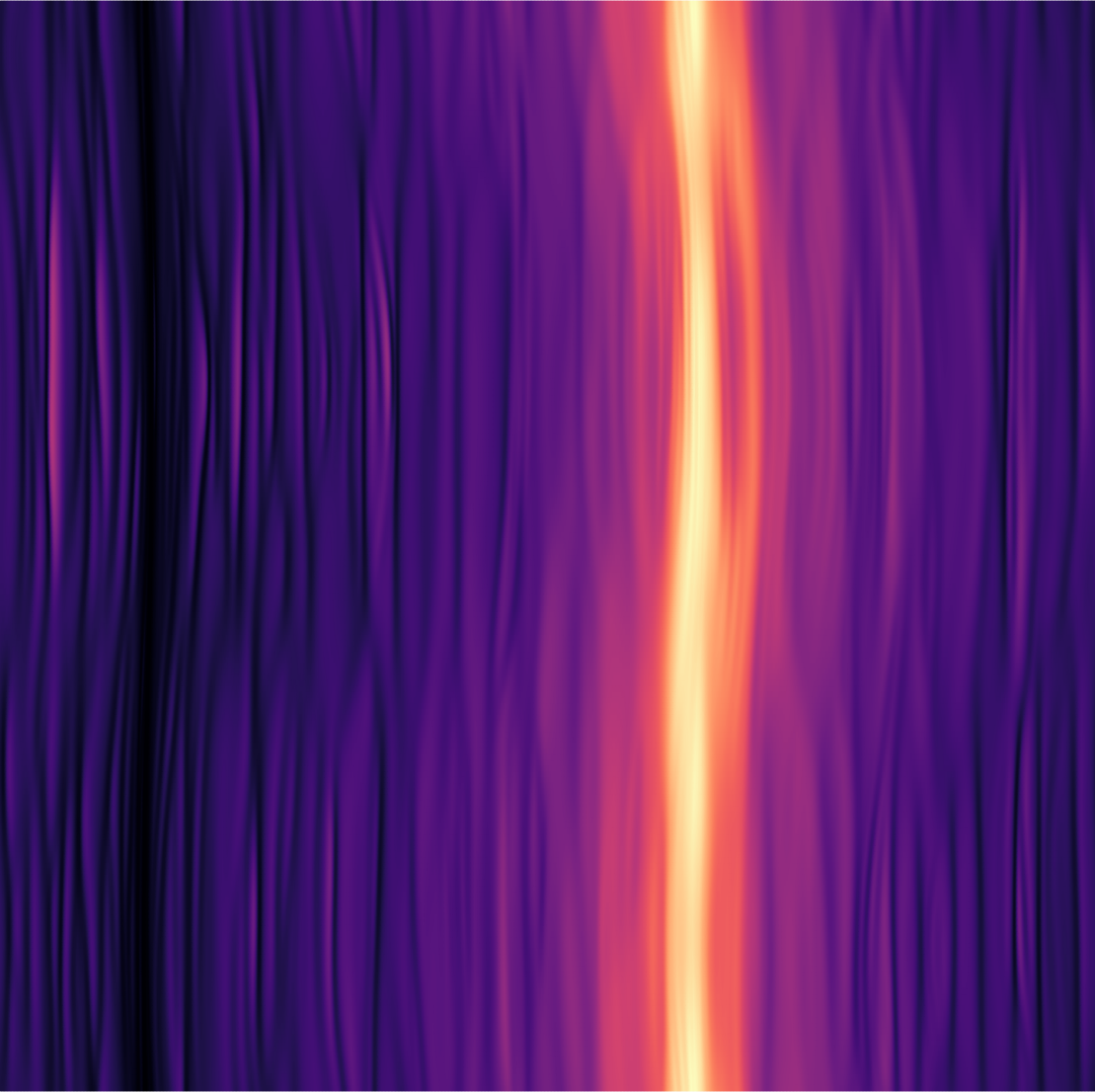}\\[1mm]
        \includegraphics[width=0.22\textwidth]{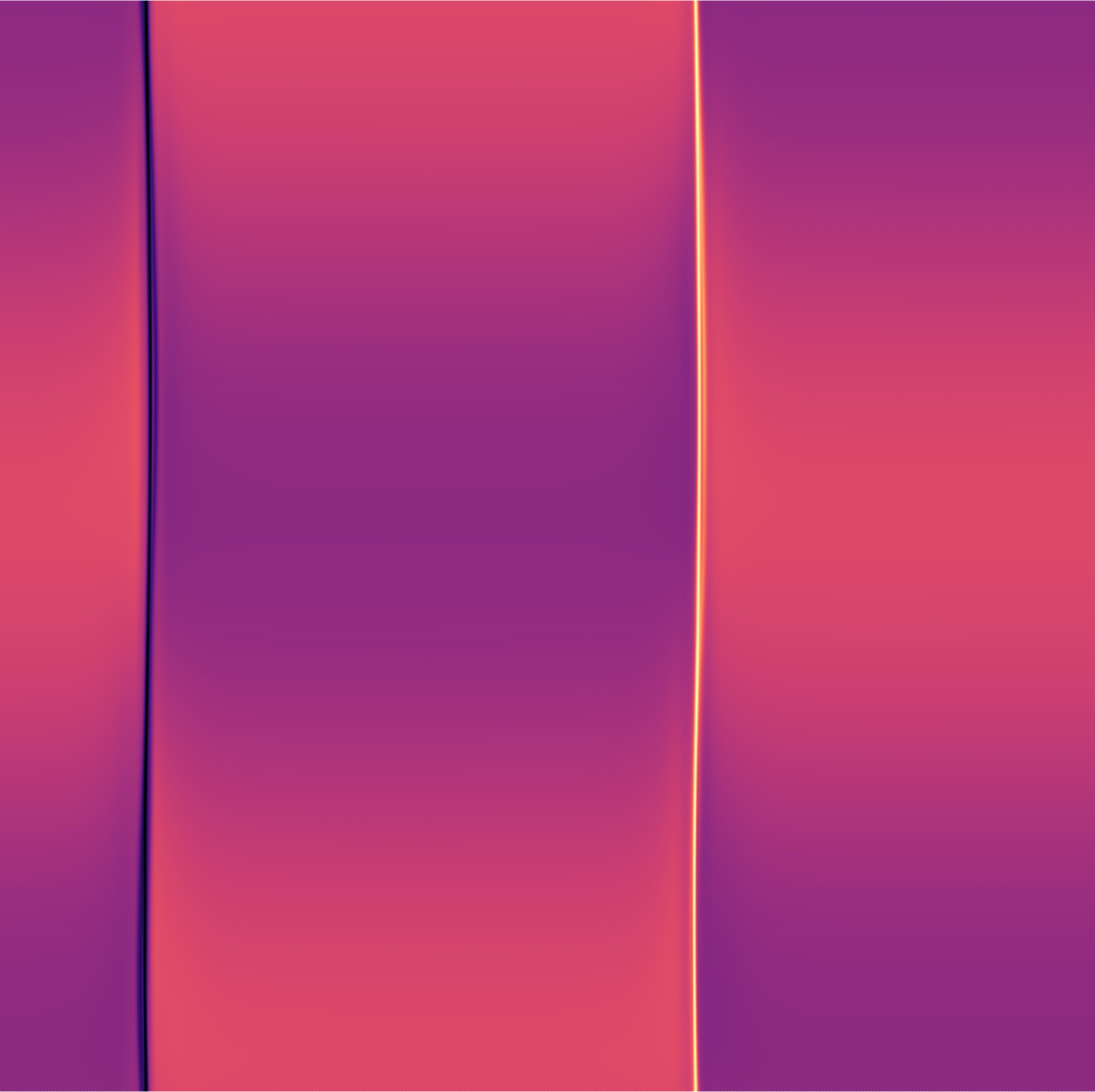}\\[1mm]
      \end{tabular}}   \\
      \subfloat[]{\label{fig:left}%
      \begin{tabular}[b]{c}
        \includegraphics[width=0.9\textwidth]{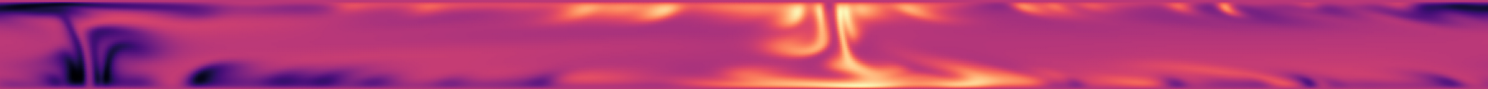}
      \end{tabular}}                \\
      \subfloat[]{\label{fig:left}%
      \begin{tabular}[b]{c}
        \includegraphics[width=0.9\textwidth]{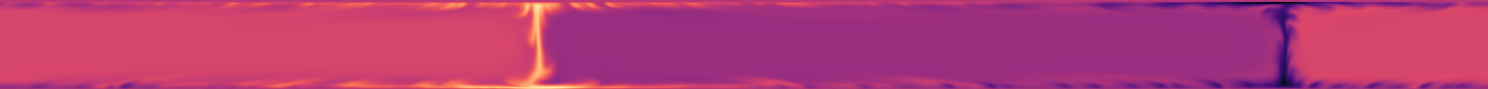}
      \end{tabular}}           
      \caption{Visualizations of the fluctuating temperature for large $Q$ cases that exhibit three-dimensional motion: (a,c) $Q=10^4$, $Ra=10^6$ and (b,d) $Q=10^6$, $Ra=1.5\times10^7$. Horizontal $x$-$y$ slices are shown in (a) and (b); vertical $y$-$z$ slices are shown in (c) and (d). The top (bottom) panels in (a) and (b) are taken at the depth of the thermal boundary layer (midplane).}
      \label{F:flow2}
\end{center}
\end{figure}

Steady 2D states persist over an increasingly larger range of $Ra$ as $Q$ is increased. Indeed, 2D states persist up to $Ra=5\times10^5$ and $Ra=2\times10^6$ for $Q=10^5$ and $Q=10^6$, respectively. To show the typical structure observed in these larger $Q$ simulations once the flow transitions to 3D, we show visualizations of the fluctuating temperature in Fig.~\ref{F:flow2} for $Q=10^4$ and $Q=10^6$. Slices through the thermal boundary layer are shown in the top panels of (a) and (b), slices through the mid-plane are shown in the bottom panels of (a) and (b), and cross-sections through the $y$-$z$ plane are shown in panels (c) and (d). For both of these simulations we find two convective rolls with a large scale modulation in the direction of the imposed magnetic field. The slices through the thermal boundary layer show the presence of small-scale anisotropic structures that are sheared by the strong $y$-directed flows that are present in the vicinity of the boundaries. The vertical slices in (c) and (d) show this tendency for shearing. Note that these cases are still within the second regime given that horizontally averaged mean flows remain dynamically negligible at these parameter values.

\subsection{Heat and momentum transport}

\begin{figure}[t!]
 \begin{center}
      \subfloat[][]{\includegraphics[width=0.48\textwidth]{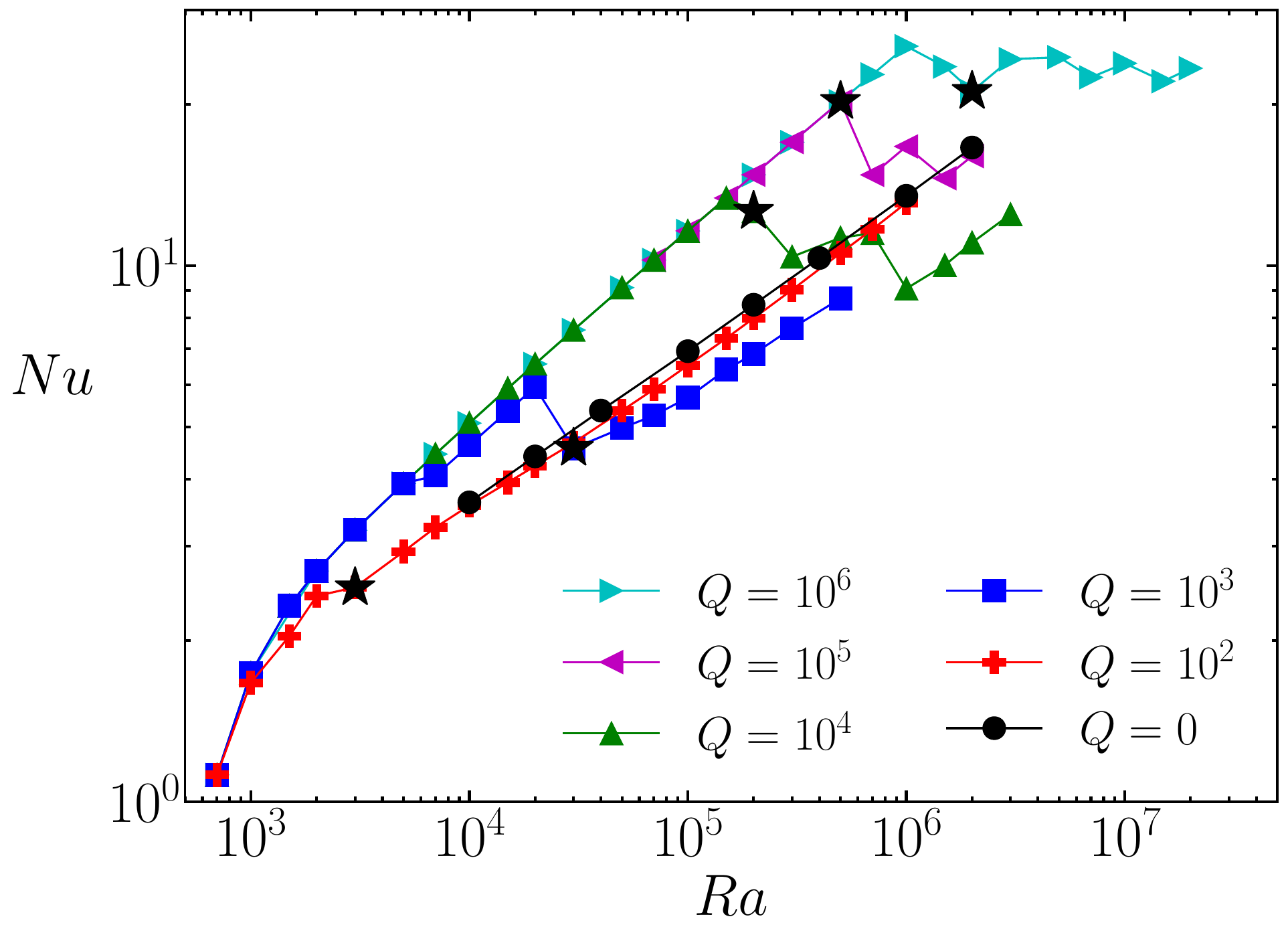}} \quad
      \subfloat[][]{\includegraphics[width=0.48\textwidth]{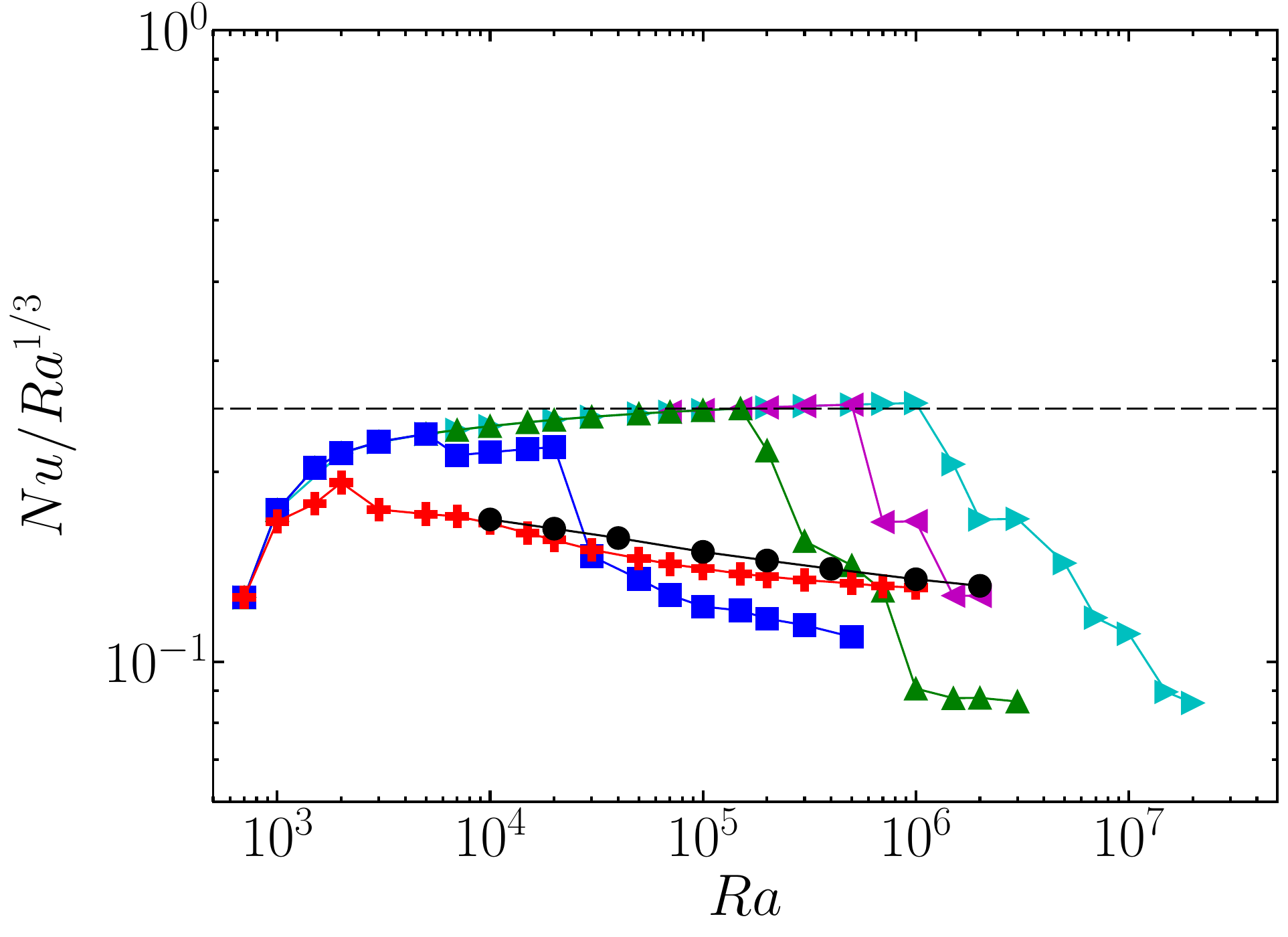}}     \\ 
      \subfloat[][]{\includegraphics[width=0.48\textwidth]{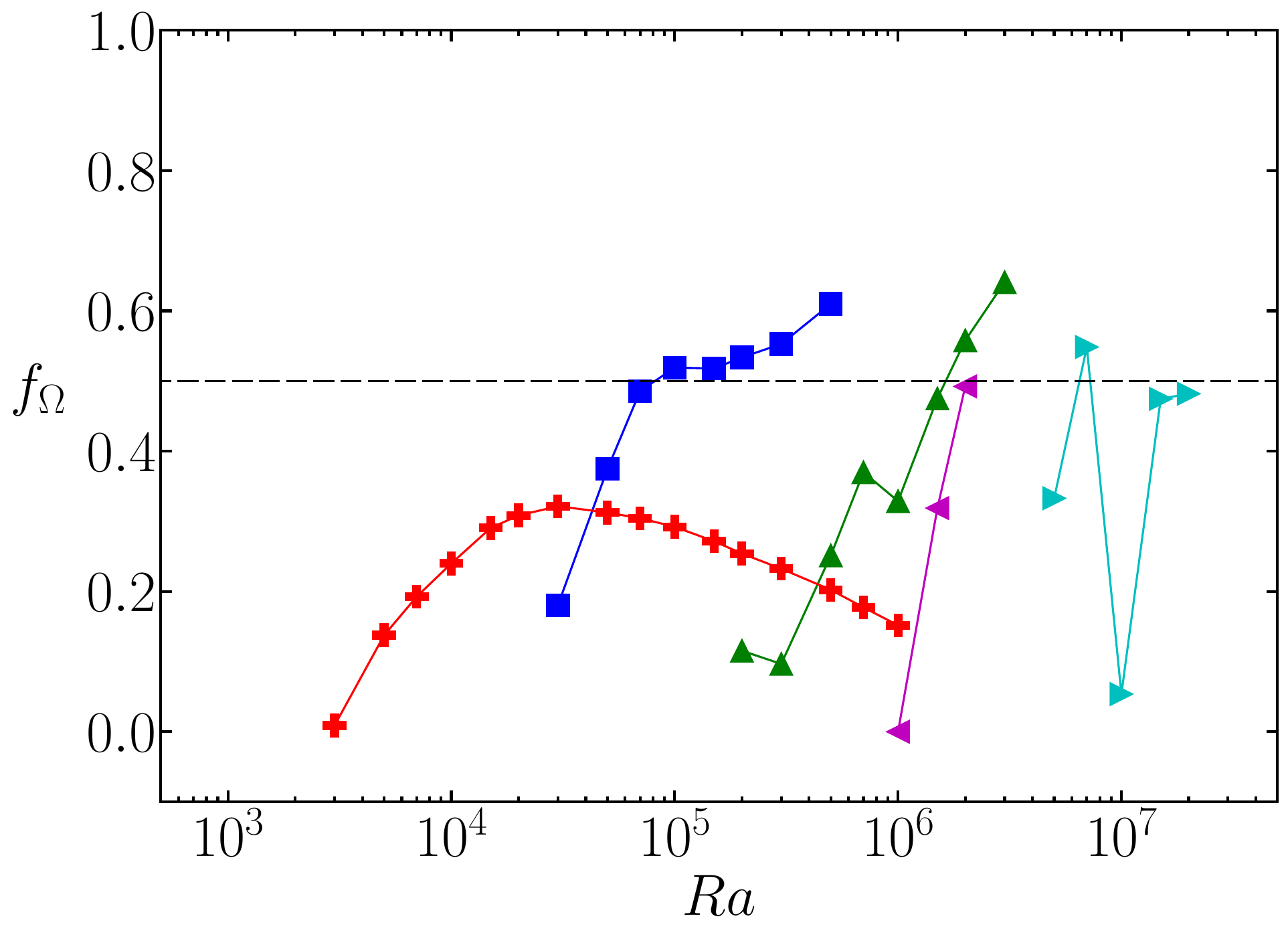}} \quad
      \subfloat[][]{\includegraphics[width=0.48\textwidth]{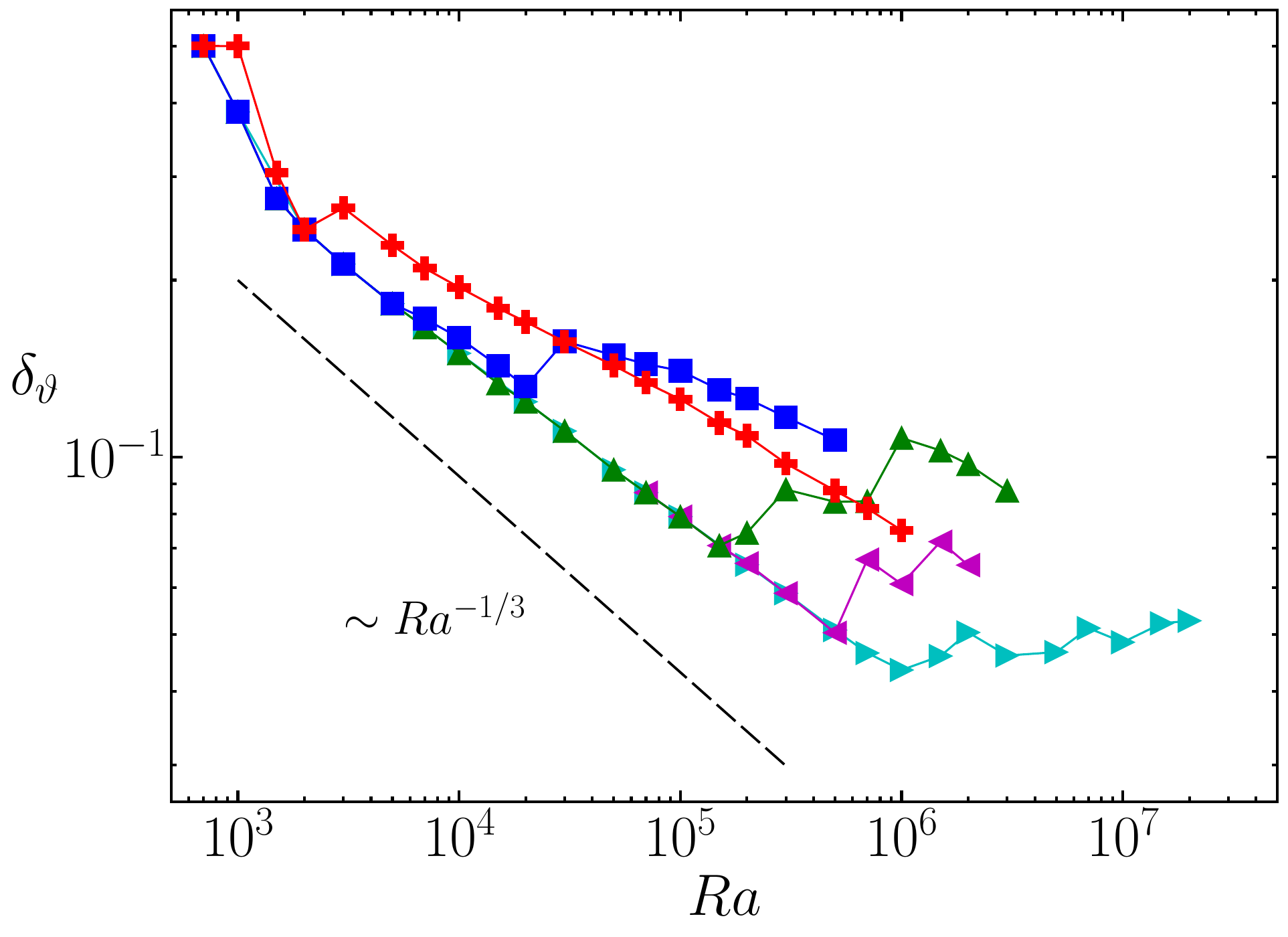}} 
      \caption{Heat transport data \textcolor{black}{versus Rayleigh number ($Ra$)} for all cases: (a) Nusselt number ($Nu$); (b) compensated Nusselt number, $Nu/Ra^{1/3}$; (c) \textcolor{black}{fraction of ohmic dissipation, $f_\Omega$}; and (d) thermal boundary layer thickness. The starred data points in (a) indicate the first observation of three-dimensional flows for the given value of $Q$. The dashed horizontal line in (b) is the prefactor computed by Chini \& Cox \cite{gC09} corresponding to wavenumber $k\approx2.22$. }
      \label{F:Nu}
\end{center}
\end{figure}

\begin{figure}[t!]
 \begin{center}
      \subfloat[][]{\includegraphics[width=0.48\textwidth]{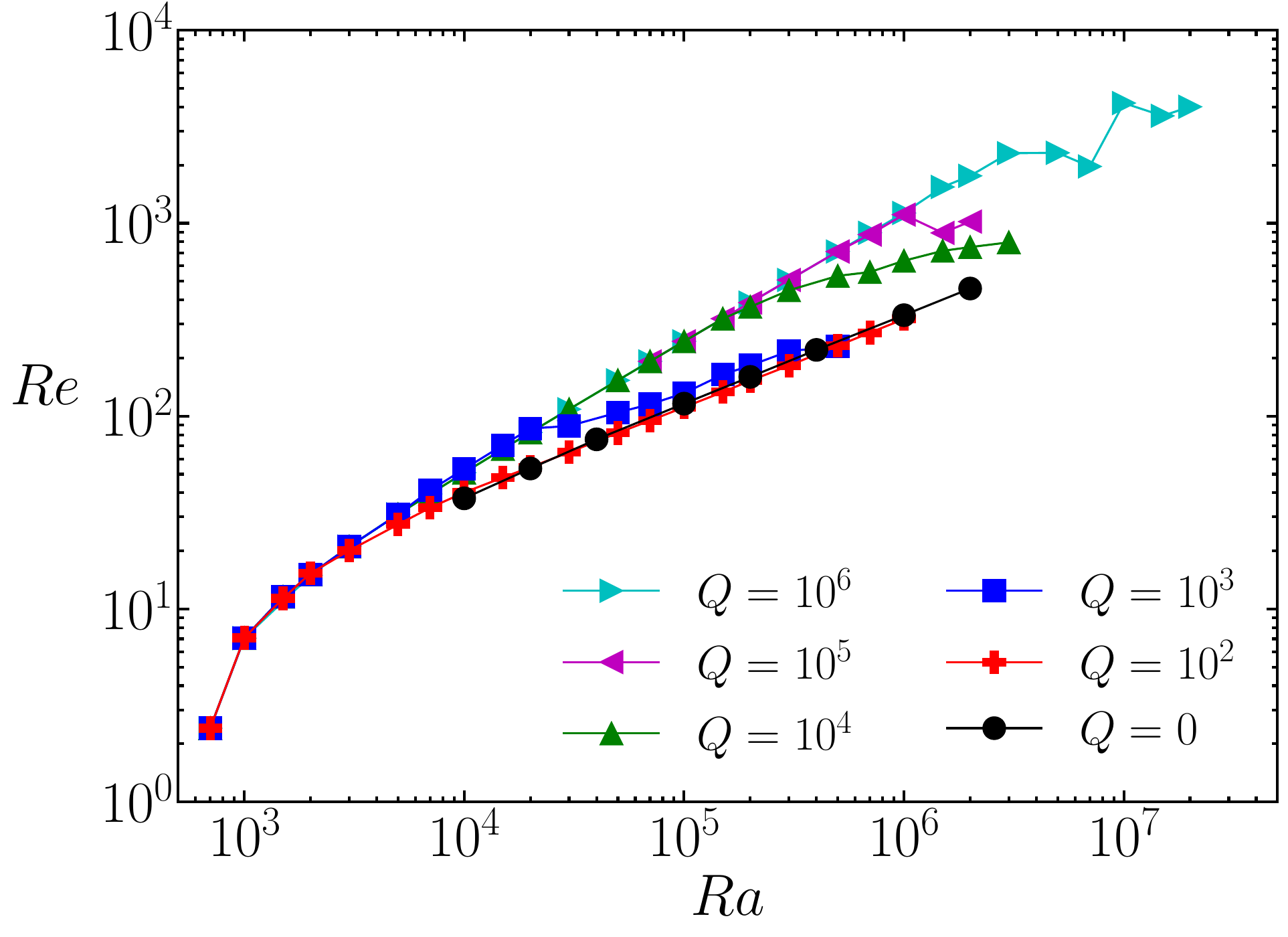}} \quad
      \subfloat[][]{\includegraphics[width=0.48\textwidth]{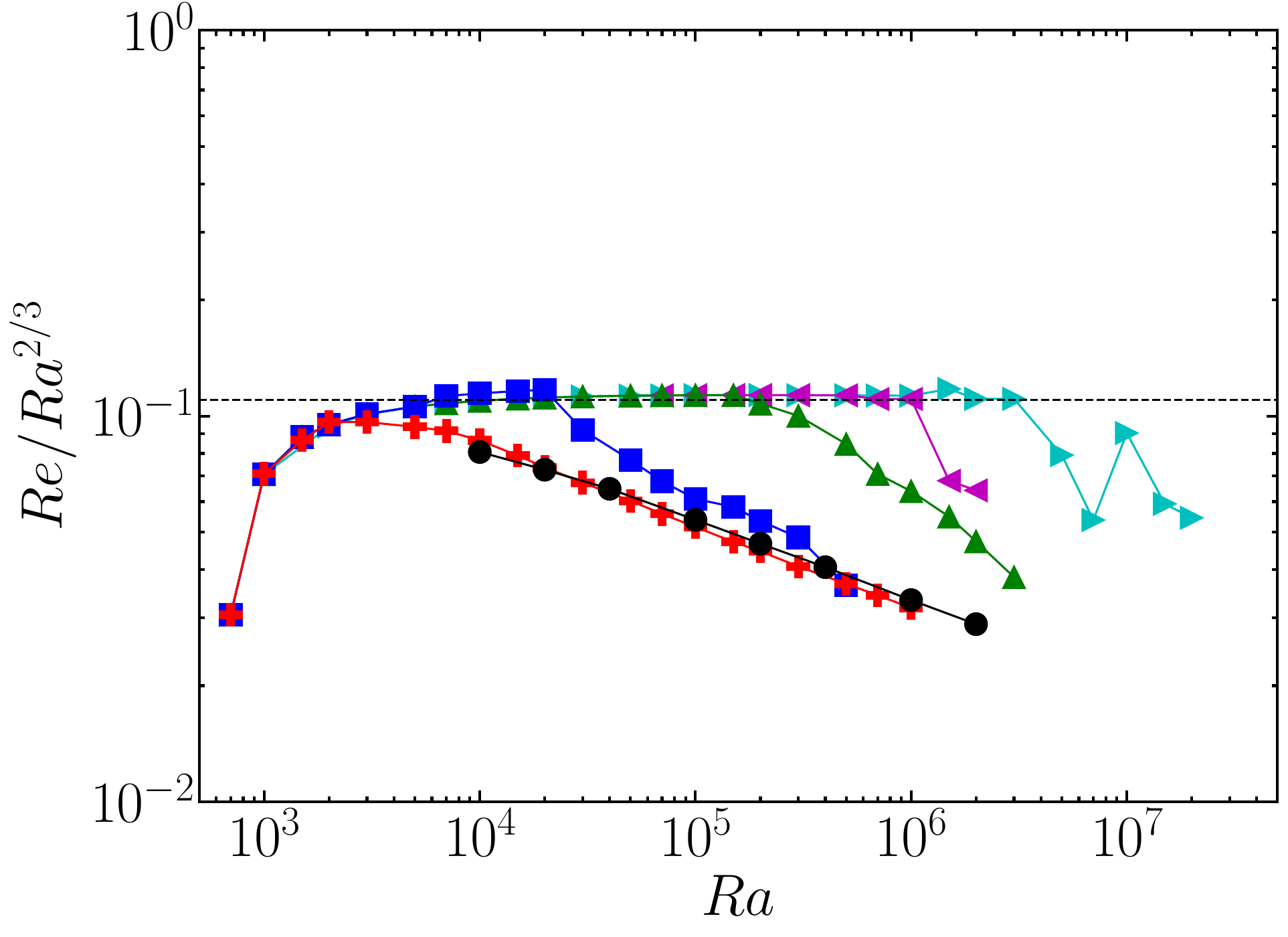}}      \\
      \subfloat[][]{\includegraphics[width=0.48\textwidth]{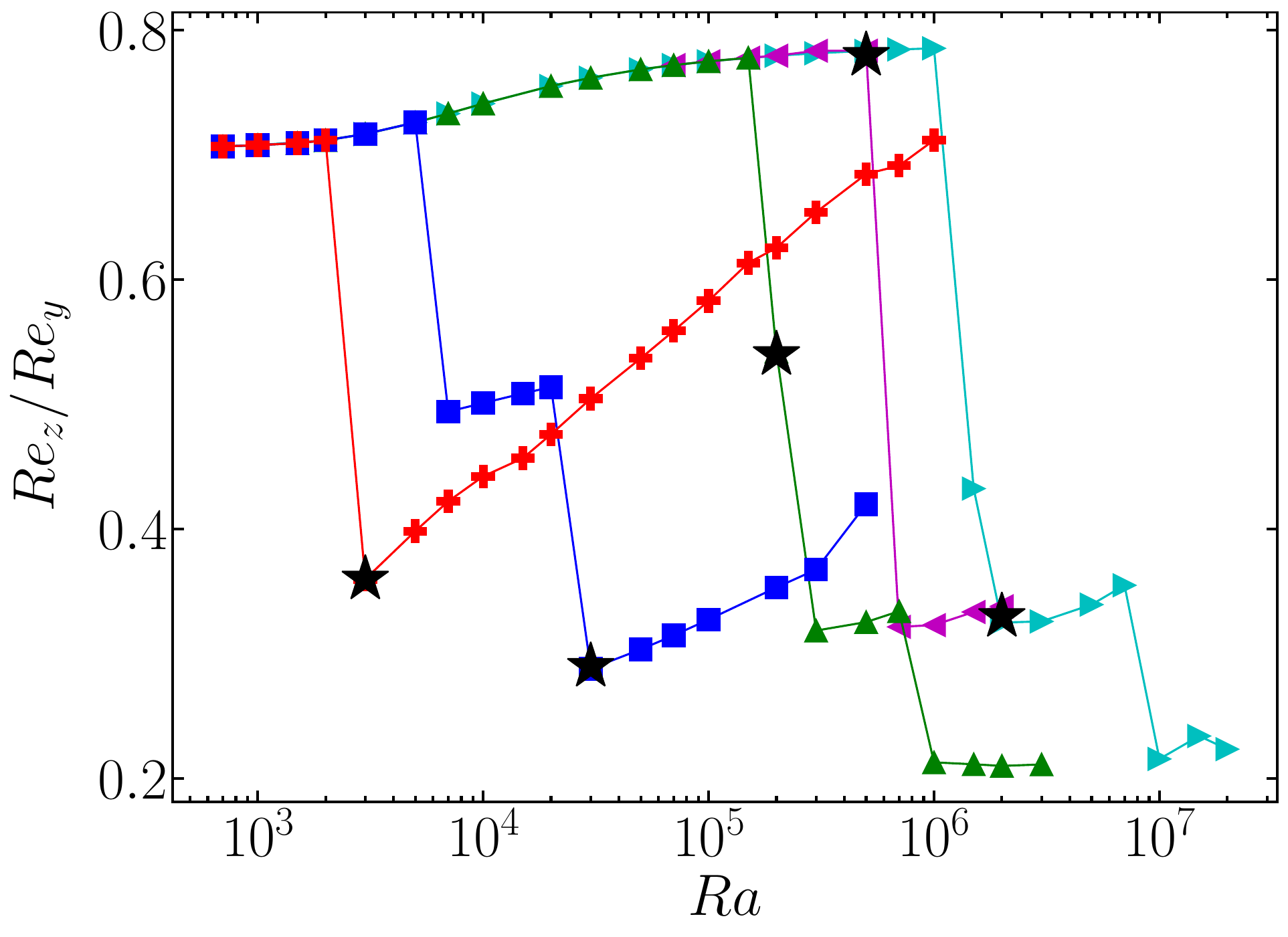}}  
      
      \caption{Momentum transport versus Rayleigh number for all cases: (a) Reynolds number, $Re$; (b) compensated Reynolds number, $Re/Ra^{2/3}$; (c) ratio of the vertical Reynolds number to transverse Reynolds number. In (b), the horizontal line at $ Re Ra^{-2/3} = 0.11$ is the asymptotic result from Ref.~\cite{bW20}. Black stars in (c) denote the values of $Ra$ for which three-dimensional motion is first observed for each value of $Q$.}
      \label{F:Re}
\end{center}
\end{figure}

To further quantify the behavior of this system, we characterize the heat and momentum transport in Figs.~\ref{F:Nu} and \ref{F:Re}, respectively. Fig.~\ref{F:Nu}(a) shows the Nusselt number, $Nu$, as a function of the Rayleigh number, $Ra$, for all simulations. The corresponding compensated values, $Nu/Ra^{1/3}$, are shown in panel (b). The dashed line in (b) is the proportionality constant computed by Ref.~\cite{gC09} for a horizontal wavenumber of $k=2.22$; this value is relevant for our simulations whenever $n_r=12$. For $Q =10^2$ the flow is 3D for $Ra \ge 3 \times 10^3$ and the heat transport scaling closely follows that of RBC, though the net heat transport is always smaller than the hydrodynamic case. 
For $Q=10^3$ and $Q=10^4$, steady 2D states with heat transport that exceeds that of RBC persist up to $Ra = 3 \times 10^4$ and $Ra = 2 \times 10^5$, respectively, and then eventually transition to heat transport scaling behavior that is also similar to RBC. Thus, at least for $Q < 10^5$, our data indicates that heat transport scaling in HMC is similar to RBC once the flow transitions to 3D, but is reduced overall. For $Q=10^6$ and $Ra > 10^6$ we find no obvious asymptotic scaling behavior for our investigated parameter range since the flow states change quickly as $Ra$ increases.


The HMC data shows several features that are consistent with the asymptotic analysis of Chini and Cox, including: (1) heat transport is most efficient when convection is 2D and the aspect ratio of convection rolls are order unity (i.e.~$n_r=12$ in our aspect ratio); and (2) all 2D states appear to approach the $Nu \sim Ra^{1/3}$ scaling at large values of $Ra$, as shown in panel (b). This latter point is particularly noticeable in the data for $Q=10^3$ shown in panel (b), which exhibits a scaling behavior over the range $7\times 10^3 \le Ra \le 3 \times 10^4$ (where $n_r=8$) that mirrors the higher $Q$ scaling directly above it (where $n_r=12$).

The fraction of ohmic dissipation is shown in Fig.~\ref{F:Nu}(c) for the 3D HMC states. For $Q=10^2$ we find that the fraction of ohmic dissipation peaks at a value of $f_{\Omega} \approx 0.32$ where $Ra = 3 \times 10^4$, then decreases monotonically, suggesting that, at least for this particular value of $Q$, HMC may approach RBC as $Ra$ is increased further. For $Q \ge 10^3$ we find that ohmic dissipation accounts for at least half of the total dissipation for $Ra \ge 10^5$. States in which $f_{\Omega} \gtrsim 0.5$ are also observed at larger $Q$. Sudden drops in $f_{\Omega}$ occur in the data for $Q=10^4$ and $Q=10^6$; these drops are associated with a reduction in the number of convection rolls.


Fig.~\ref{F:Nu}(d) shows the thermal boundary layer thickness $\delta_\vartheta$, defined as the distance between the horizontal boundary and the location of the maximum value of the time-averaged rms of $\vartheta'$. When the heat transport is limited by conduction across the thermal boundary layer, we expect that the Nusselt number scales as
\be
Nu \sim \delta_\vartheta^{-1}.
\ee
We therefore expect $\delta_\vartheta \sim Ra^{-1/3}$ whenever the $Nu \sim Ra^{1/3}$ scaling holds -- this slope is indicated in the figure. The thermal boundary thickness scaling is clearly correlated with the scaling of $Nu$ in the sense that enhancement (reduction) of heat transport is associated with a decrease (increase) in the thermal boundary layer thickness. For $Q = 10^6$, the thermal boundary layer thickness shows a slight overall increase with Rayleigh number for $Ra > 10^6$.



The rms convective flow speeds, as characterized by the Reynolds number, are shown in Fig.~\ref{F:Re}(a). Compensated values of the Reynolds number, $Re/Ra^{2/3}$, are shown in Fig.~\ref{F:Re}(b). This $Re \sim Ra^{2/3}$ scaling is consistent with the theory of Ref.~\cite{gC09}, though the Reynolds number is only explicitly discussed in Ref.~\cite{bW20}. The horizontal line in (b) is the asymptotic value taken from Ref.~\cite{bW20} for a horizontal wavenumber of $k \approx 2.22$, which characterizes our roll state with $n_r=12$. As for the heat transport data, our magnetically constrained 2D states appear to closely follow the Chini \& Cox scaling predictions. Once the flow becomes 3D we find reductions in the efficiency of momentum transport, as indicated by departures from the $Re \sim Ra^{2/3}$ scaling.

The ratio of the vertical ($z$) Reynolds number to the transverse ($y$) Reynolds number is given in Fig.~\ref{F:Re}(c). From mass conservation we expect this ratio to be correlated with $\Gamma_r$ -- larger values of $Re_z/Re_y$ typically coincide with smaller values of $\Gamma_r$ and therefore, in accordance with Fig.~\ref{F:cells}(b), with smaller values of $Ra$. There is a clear `optimal' (in the sense of heat and momentum transport) trend near the top of the plot where all cases are 2D. However, some values of $Q$ show a departure from this trend even when the flow is 2D. For example, for $Q=10^3$ we find an intermediate regime in the vicinity of $Ra = 10^4$, where the scaling of the Reynolds number ratio is similar to the `optimal' curve. The Reynolds number ratio for $Q=10^2$ exhibits a monotonic increase with $Ra$ and shows a steeper slope in comparison to the larger $Q$ cases. Moreover, the ratio appears to be approaching unity for $Q=10^2$ even though the roll aspect ratio remains fixed at $\Gamma_r \approx 3$. The three largest values of $Q$ show similar behavior in which there is a transition to $Re_z/Re_y \approx 0.3$, and then a transition to $Re_z/Re_y \approx 0.2$ at the largest accessible values of $Ra$ in which only two convection rolls are present.

\subsection{Transition to 3D convection}

\begin{figure}
 \begin{center}
\subfloat[][]{\includegraphics[width=0.48\textwidth]{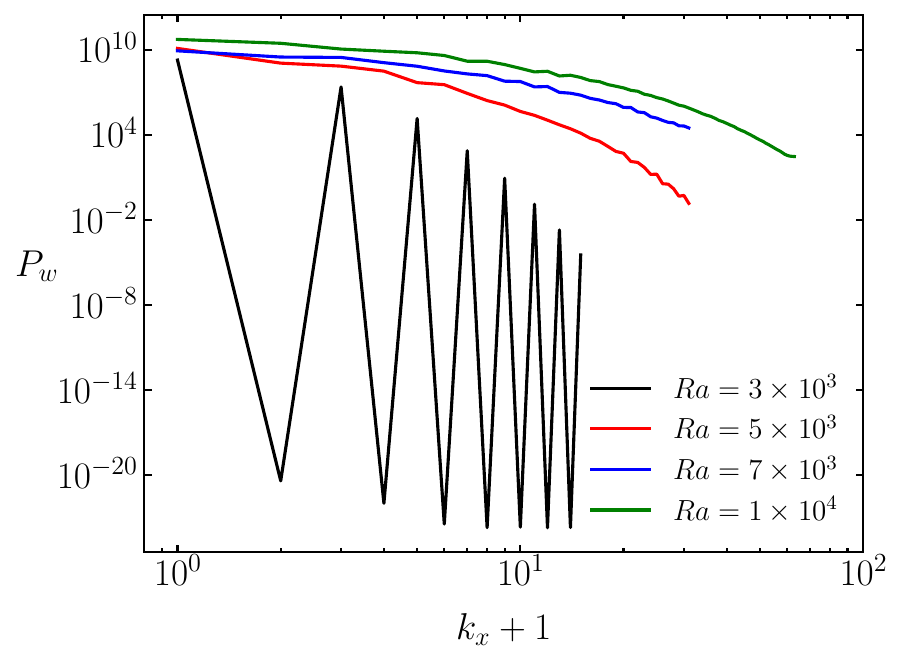}}  \quad
\subfloat[][]{\includegraphics[width=0.48\textwidth]{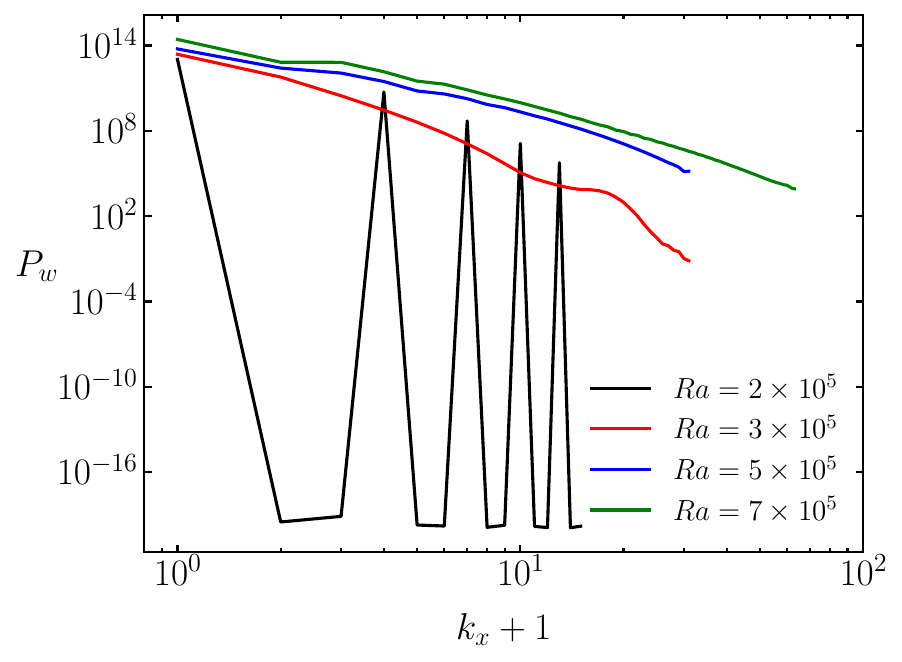}} \\
\subfloat[][]{\includegraphics[width=0.48\textwidth]{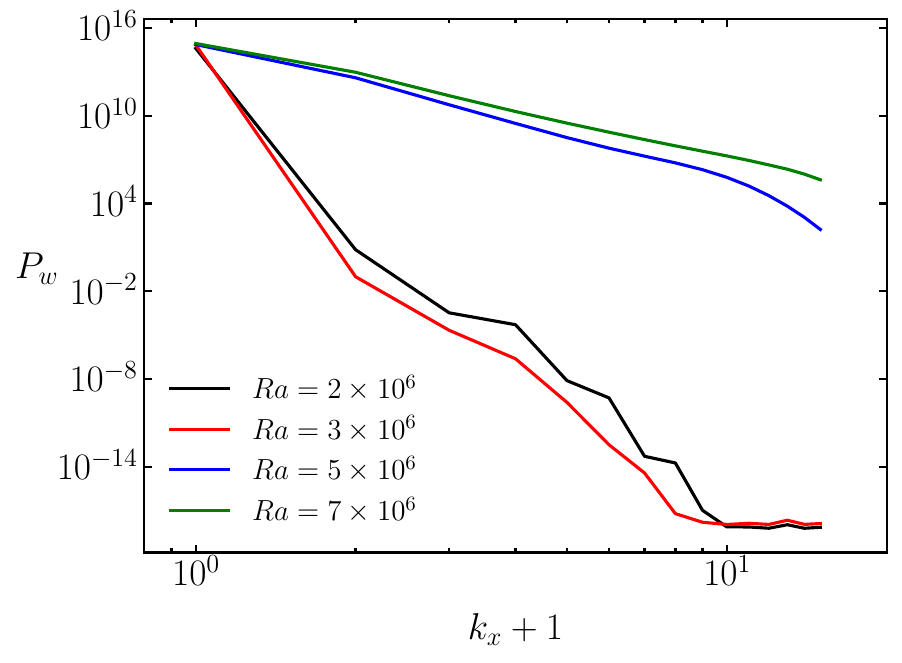}}
      \caption{Instantaneous one dimensional (along the direction of the magnetic field) power spectra of the vertical component of the velocity vector \textcolor{black}{showing the onset of three-dimensional motion}: (a) $Q = 10^2$; (b) $Q = 10^4$; (c) $Q = 10^6$. The spectra are \textcolor{black}{averaged over the depth of the layer}. \textcolor{black}{For each value of $Q$, the smallest value of $Ra$ shown corresponds to the first observed onset of three-dimensional motion.}}
      \label{F:wspectral}
\end{center}
\end{figure}

Cases that are magnetically constrained (i.e.~2D) exhibit a relatively sudden transition from steady 2D convection to 3D convection as $Ra$ is increased. In terms of the horizontal wavenumber $\kb = (k_x, k_y)$, this transition coincides with an appearance of flows in which $k_x \ne 0$. Thus, Fourier spectra can be used to characterize the transition. For this purpose we define the \textcolor{black}{depth-averaged} one dimensional power spectrum for the vertical velocity component,
\be
\textcolor{black}{P_w(k_x) = \frac{1}{2} \int_0^{1} \sum_{k_y} \lsq \widehat{w}(k_x, k_y, z) \cdot \widehat{w}^{*}(k_x, k_y, z)  \rsq dz,} \label{2DFFT}
\ee
where $\widehat{w}(k_x, k_y, z)$ is the Fourier transform of the vertical velocity at a particular height and the star superscript denotes a complex conjugate. 
We refer to convection as 3D if the computed power present in modes with $k_x \ne 0$ is above machine precision.
In Fig.~\ref{F:wspectral} this power spectrum is shown for instantaneous data for $Q = (10^2, 10^4, 10^6)$ and \textcolor{black}{four different values of the Rayleigh number, where the smallest value of $Ra$ for each value of $Q$ corresponds to the first observed onset of 3D motion}. For $Q = 10^2$, \textcolor{black}{the onset of 3D motion occurs at $Ra=3\times10^3$ and Fig.~\ref{F:wspectral}(a) indicates that this 3D flow is dominated by a $k_x = 2$ mode with higher harmonics generated through nonlinear interactions. At $Ra=3\times10^3$ the convection is steady, and the transition to unsteady broadband convection occurs at $Ra=5\times10^3$ for this value of $Q$. However, the spectra for $Ra = 7\times10^3$ and $Ra =10^4$ show that the $k_x=0$ mode remains dominant, and this mode is observed to remain dominant up to the largest value considered for this value of $Q$ ($Ra=10^6$).}

Fig.~\ref{F:wspectral}(b) shows spectra for $Q=10^4$, where the first 3D convection occurs at $Ra=2\times10^5$ and is dominated by a $k_x=3$ mode. We find unsteady 3D convection for $Ra > 3\times10^5$. At $Ra=3\times10^5$ the dominant 3D mode is $k_x=1$ though the spectra are broadband. For $Q=10^6$, Figure \ref{F:wspectral}(c) shows that the first 3D mode occurs at $Ra=2\times10^6$ and consists of a $k_x=1$ structure and this mode remains the dominant 3D mode for all higher Rayleigh numbers. Note that the energy present in $k_x > 0$ modes decreases when the Rayleigh number is increased from $Ra=2\times10^6$ to $Ra=3\times10^6$. For $Q=10^6$ we first observe unsteady convection at $Ra = 7\times10^6$ -- the other three Rayleigh number cases shown in panel (c) are all steady in time. We find that the 2D mode, characterized by $k_x=0$ remains dominant for all parameter values investigated, indicating that anisotropy remains persistent.

\begin{figure}
 \begin{center}
\includegraphics[width=0.5\textwidth]{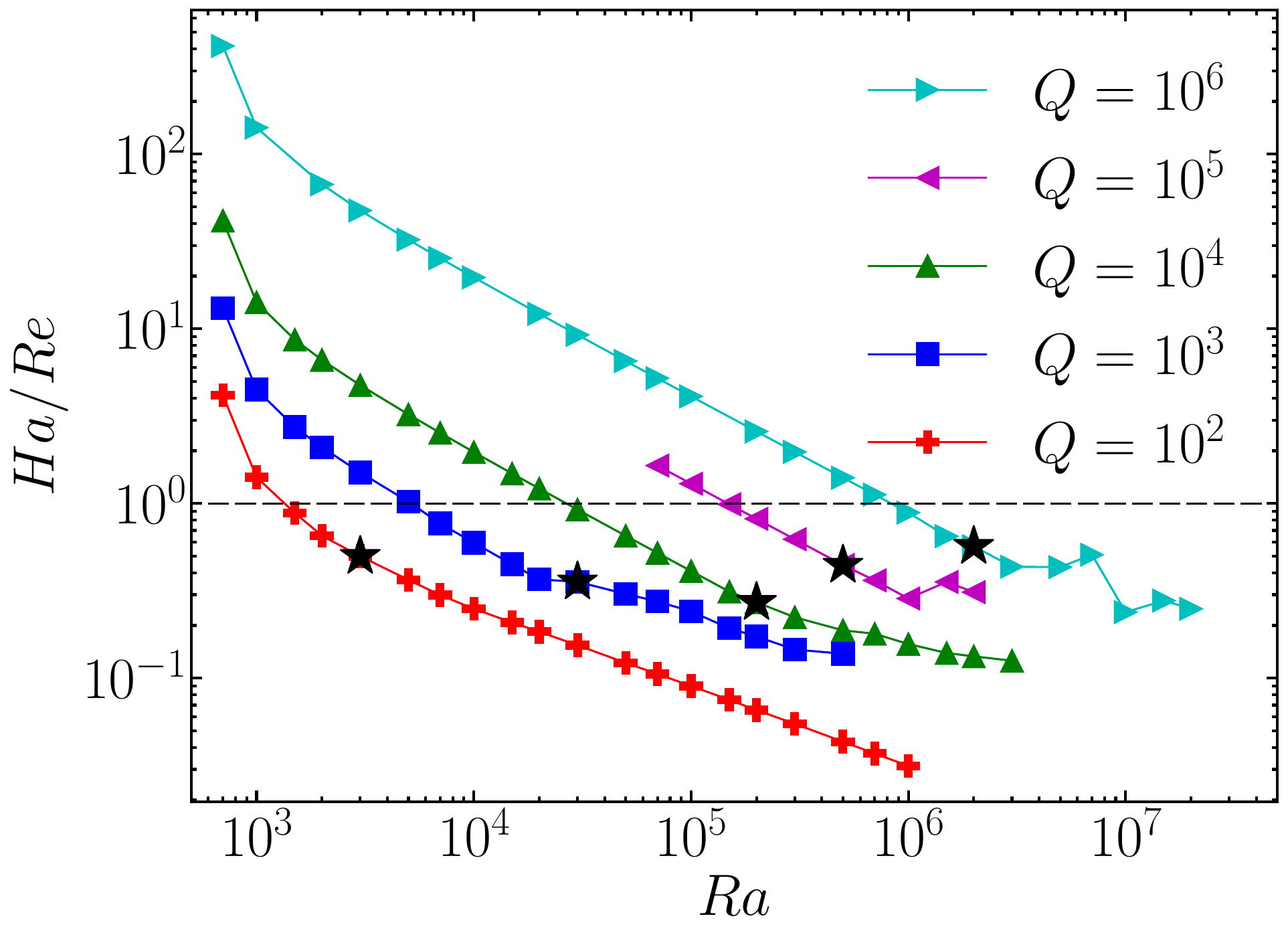}
      \caption{Characterization of the transition from 2D to 3D convection with the ratio $Ha/Re = \sqrt{Q}/Re$. Starred symbols denote the smallest value of $Ra$ for which 3D motion is observed for the indicated value of $Q$.} 
      \label{F:HaRe}
\end{center}
\end{figure}

The Lorentz force acts to suppress variations of the flow field along the direction of the imposed magnetic field. We expect this transition to occur at larger values of $Ra$ as $Q$ is increased since larger flow speeds are necessary to overcome the constraint imposed by the magnetic field. Based on this observation, a simple argument for estimating the transition to 3D states consists of comparing the Lorentz and inertial forces. 
In the quasi-static limit used in the present investigation, the ratio of the Lorentz force to the viscous force is characterized by the Hartmann number \citep[e.g.][]{hunt1971magnetohydrodynamics,bK08},
\be
Ha=\frac{B_{0} H}{\sqrt{\rho \nu \mu \eta}} = \sqrt{Q} .
\ee
Thus, the ratio $Ha/Re$ represents the relative size of the Lorentz force and inertial forces in the present study. Fig.~\ref{F:HaRe} shows this ratio for all of the simulations. The smallest value of $Ra$ for which 3D motion is first observed is denoted by the starred data points for each value of $Q$. For reference, the dashed horizontal line shows $Ha/Re=1$. We observe some scatter in the value of $Ha/Re$ for which 3D motion is first observed for the different values of $Q$; for $Q=(10^2, 10^3, 10^4, 10^5, 10^6)$ we have  $Ha/Re = (0.50, 0.36, 0.27, 0.44, 0.57)$, respectively, for the transitional values. Nevertheless, all 2D/3D transitional values of $Ha/Re$ are of the same order of magnitude, indicating its relevance for characterizing this phenomenon. The scatter in values could be related to the structure of the eventual 3D flow; as shown in Fig.~\ref{F:wspectral} we find different $x$-dependent structure for the different values of $Q$. In this sense the ratio $Ha/Re$ may underestimate the size of the Lorentz force for cases in which the 3D state is characterized by $k_x >1$.

\subsection{Mean Flows}

\begin{figure}
 \begin{center}
      \subfloat[][]{\includegraphics[width=0.48\textwidth]{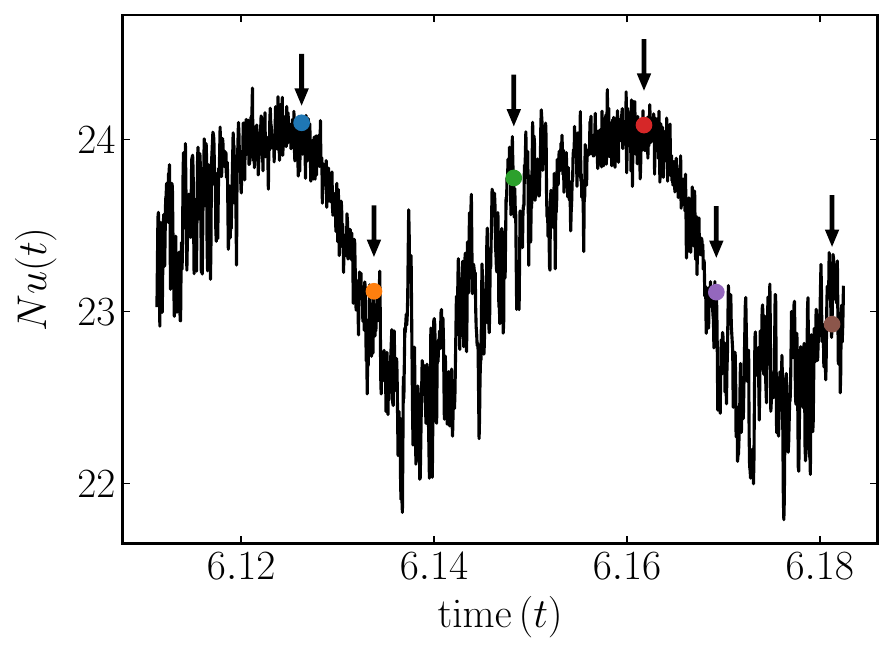}} \quad
          \subfloat[][]{\includegraphics[width=0.48\textwidth]{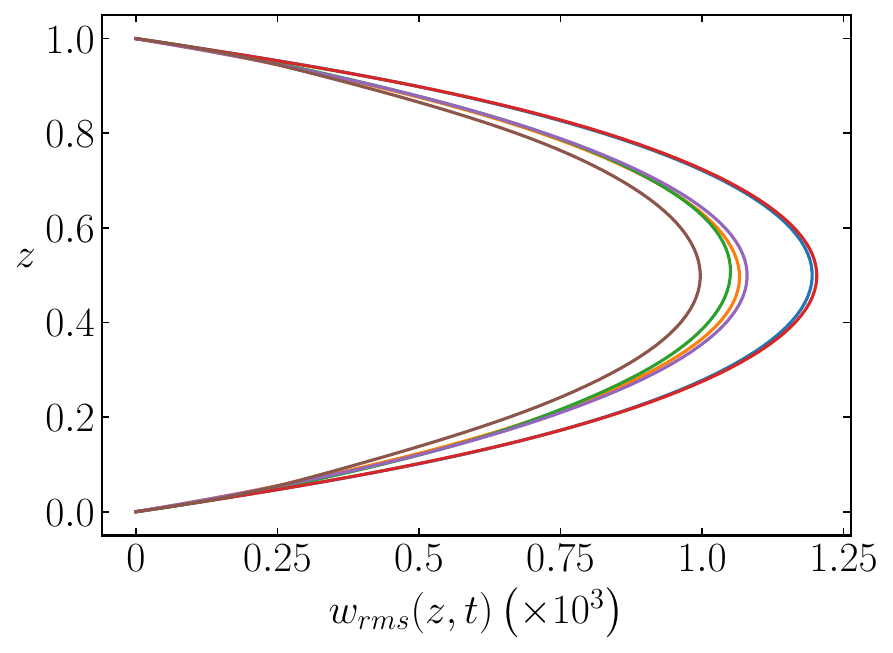}} \\
       \subfloat[][]{\includegraphics[width=0.48\textwidth]{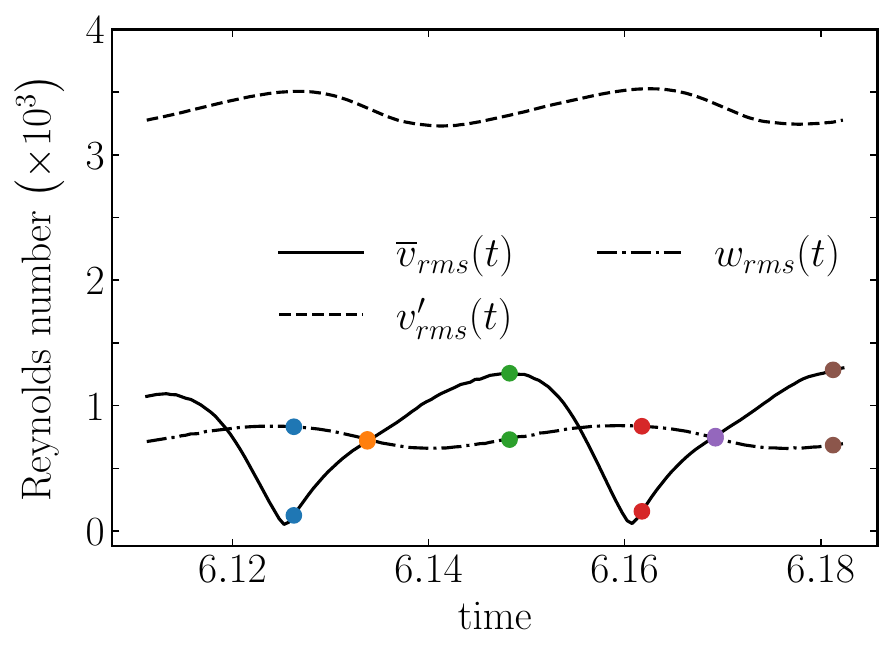}}  \quad
      \subfloat[][]{\includegraphics[width=0.48\textwidth]{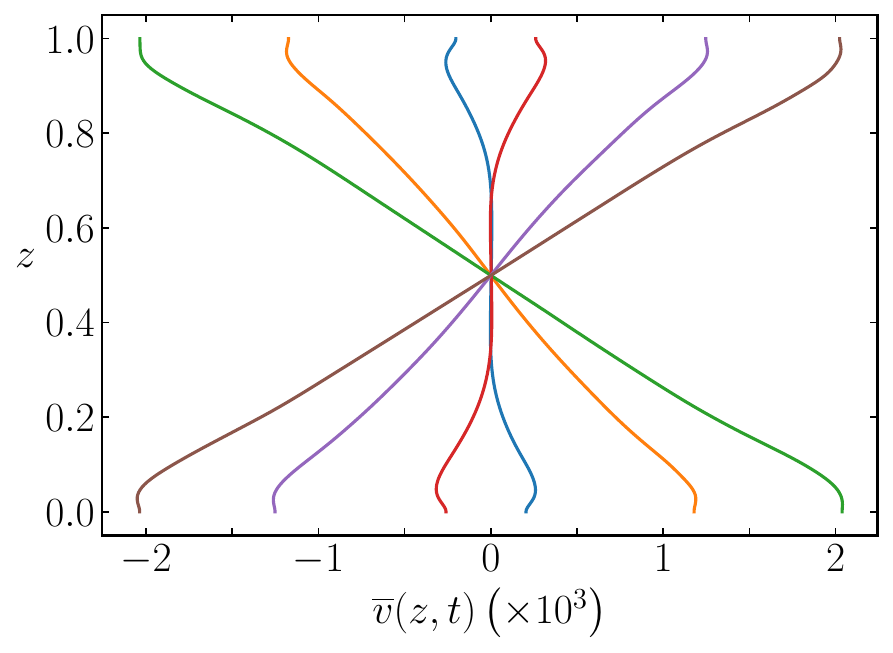}}  
      \caption{Heat transport and flow speeds from the first observed case that shows a substantial mean flow ($Q = 10^6$ and $Ra = 2 \times 10^7$): (a) time series of the Nusselt number;  (b) vertical profiles of the rms vertical velocity; (c) time series of various measures of flow speed; (d) vertical profiles of the $y$-component of the horizontally averaged flow. Dashed vertical lines in (a) and colored dots in (a) and (c) delineate the times corresponding to the profiles shown in (b) and (d).  }
      \label{F:meanflow}
\end{center}
\end{figure}

We observe the formation of strong large scale horizontal mean flows for cases with $Q=10^6$ and $Ra \ge 2 \times 10^7$. Time-dependent, strongly constrained flows appear to be a necessary ingredient for driving such flows, and it is therefore not surprising that we find them for only the most extreme simulations that were performed. Here we report on the behavior of the two cases ($Ra = 2 \times 10^7$ and $Ra = 3 \times 10^7$) for which such states were observed. In the present context we use the term `mean' to refer to averages over the horizontal plane, and recall that such quantities are denoted with an overline. The observed mean flows are characterized by a velocity that is predominantly aligned with the $y$-direction, and shows a nearly linear profile over the depth ($z$) of the layer; these flows are essentially equivalent in structure and dynamics to those found in both 2D convection \citep{dG14,qW20} and 3D convection with a horizontal rotation vector \citep{von2015generation,qW20,kL22}. However, one of the primary differences between these previous investigations that find mean flows and the present work is that the Lorentz force results in a source of dissipation and it may therefore play a different role in comparison to studies using a horizontal rotation vector since the Coriolis force is conservative. In addition, there is no preference for mean flow direction in the present case since the governing equations are invariant to the transformation $\Bb_0 \rightarrow -\Bb_0$. 

Fig.~\ref{F:meanflow} shows data from the first observed case that generates a mean flow with a magnitude that is comparable to the underlying convective flow speeds. This case is characterized by periodic dynamics in time. A time history of the instantaneous Nusselt number, $Nu(t)$, is shown in Fig.~\ref{F:meanflow}(a), and several measures of flow speeds are shown in panel (c). The periodic dynamics are evident in all of these quantities. For the Nusselt number we also observe random temporal fluctuations that are due to the small scale anisotropic structures that are localized near the top and bottom boundaries; similar structures are observed in the rotating case studied by Ref.~\citep{qW20}. The strong vertical shear that is present away from the boundaries likely prevents these structures from persisting within the fluid interior. There is a clear correlation between larger values of $Nu(t)$ and larger values of $w_{rms}(z,t)$, as shown in panels (a) and (b). Fig.~\ref{F:meanflow}(c) shows vertical averages for various rms flow speeds -- here we find that the mean flow is comparable in magnitude to the vertical component of the velocity, whereas the $y$-component of the fluctuating velocity remains dominant for this particular case. However, the vertical profiles of the mean flow shown in Fig.~\ref{F:meanflow}(d) shows that the mean flow reaches maximum values that exceed the maximum values of the (rms) vertical velocity given in panel (b). We also find a phase lag between the maximum values of the rms convective flow speeds and the maximum value of the mean flow. Since nonlinear correlations in the convection are the sole source of the mean flow, it is expected that as the convection becomes sheared out by the mean flow, an eventual decay of the mean flow will occur in time.  The mean flow varies approximately linearly with depth at times when it reaches its maximum amplitude, as indicated by the green and brown curves in panel (d).

\begin{figure}
 \begin{center}
      \subfloat[][]{\includegraphics[width=0.48\textwidth]{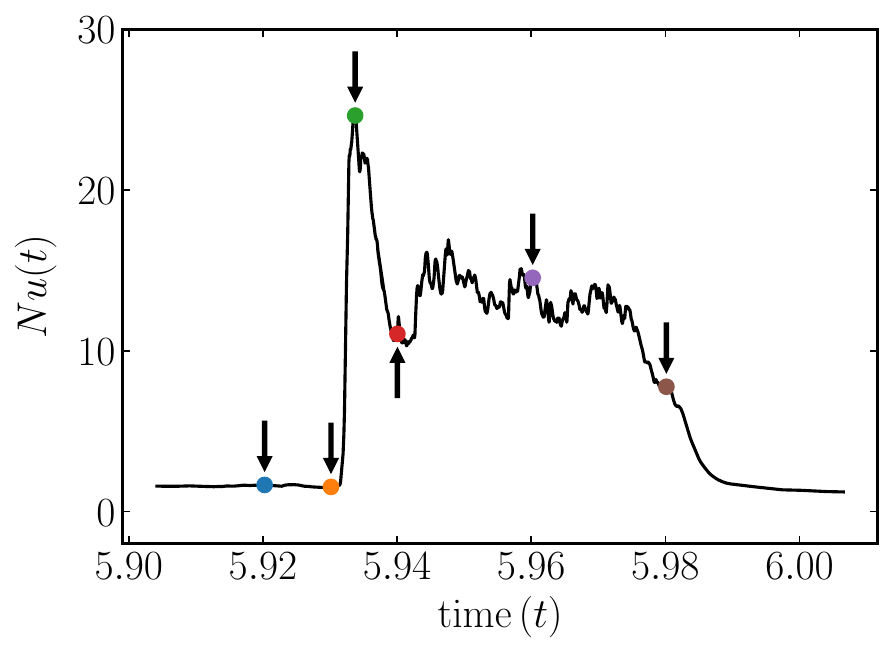}} \quad
          \subfloat[][]{\includegraphics[width=0.48\textwidth]{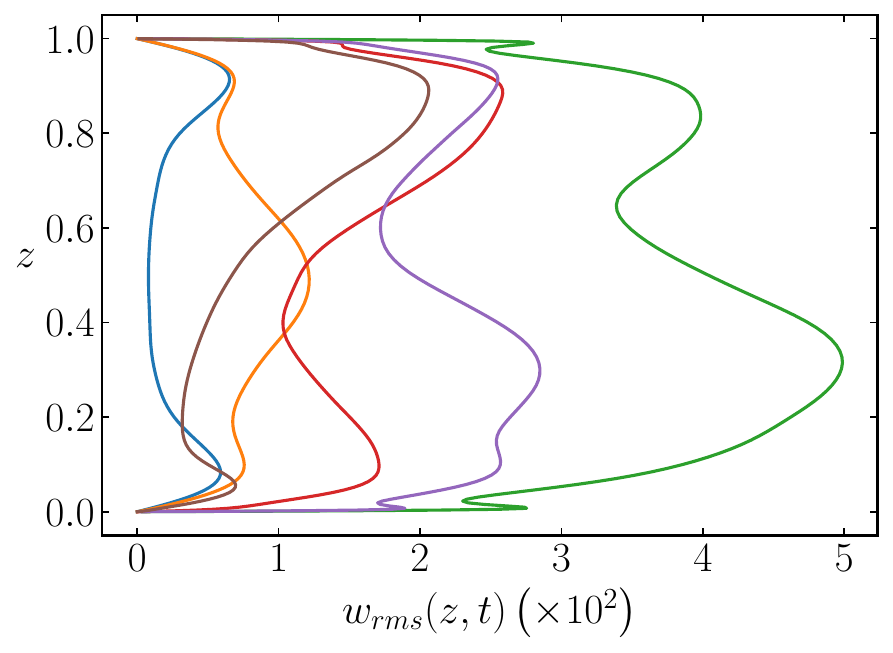}} \\
       \subfloat[][]{\includegraphics[width=0.48\textwidth]{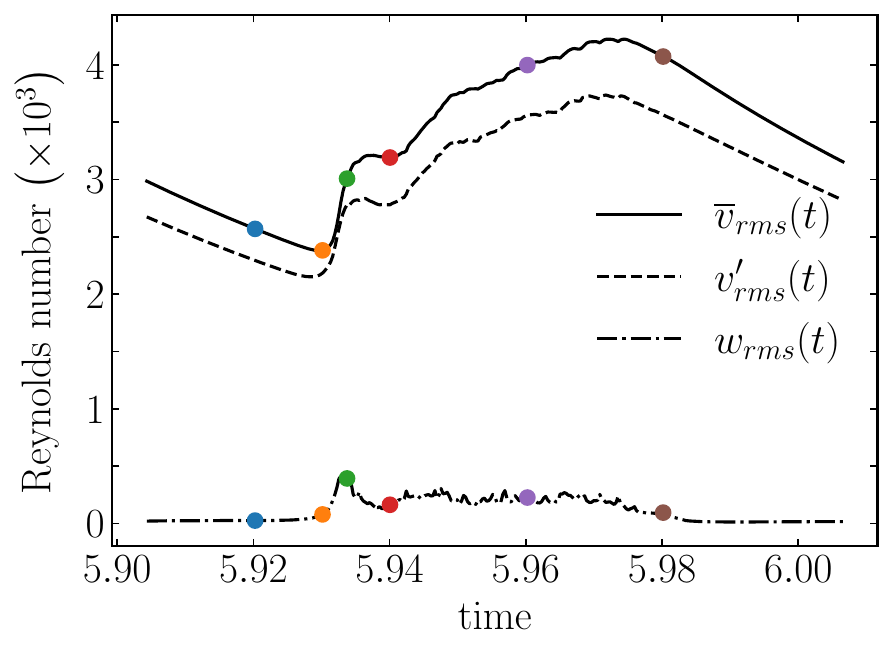}}  \quad
      \subfloat[][]{\includegraphics[width=0.48\textwidth]{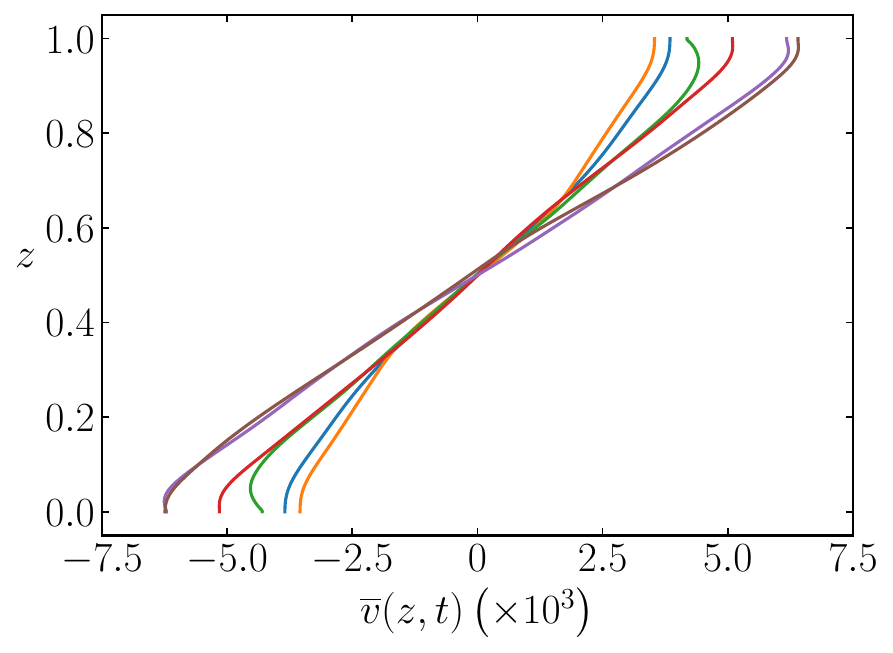}}  
      \caption{Heat transport and flow speeds from the second observed case that shows a substantial mean flow with relaxation oscillations ($Q = 10^6$ and $Ra = 3 \times 10^7$): (a) time series of the Nusselt number;  (b) vertical profiles of the rms vertical velocity; (c) time series of various measures of flow speed; (d) vertical profiles of the $y$-component of the horizontally averaged flow. Dashed vertical lines in (a) and colored dots in (a) and (c) delineate the times corresponding to the profiles shown in (b) and (d).  }
      \label{F:relax}
\end{center}
\end{figure}

As the Rayleigh number is increased to $Ra=3\times10^7$, we find a transition to a state that is characterized by relaxation oscillations with mean flows that are larger in magnitude than than the convective flow speeds [i.e.~$|\overline{v}| > (|v^{\prime}_{rms}|, |w_{rms}|$)]. Fig.~\ref{F:relax} shows output from this case. As panel (a) shows, during times in which the mean flow is strong we find average Nusselt numbers as small as $Nu \approx 1.5$, whereas $Nu \approx 12$ when the mean flow is weak, with peaks as large as $Nu \approx 25$. During a relaxation oscillation cycle, the vertical component of the velocity is as small as $\langle w_{rms} \rangle = O(10)$, whereas it reaches max values of $O(500)$ at certain points of the cycle, as shown in Figs.~\ref{F:relax}(b) and (c). Interestingly, in comparison to $Ra = 2\times10^7$, we find that the magnitude of the vertical velocity is approximately halved and that the vertical structure is considerably different, even showing the presence of a boundary layer at certain times. Panels (c) and (d) show the temporal behavior and spatial structure of the mean velocity, where we find that the magnitude of $\overline{v}$ grows during the phase of the cycle in which the heat transport is large and vice versa.

\section{Conclusions}
\label{S:conclusions}


Within the parameter space investigated, numerical simulations of magnetoconvection (MC) in a periodic plane layer geometry with an imposed horizontal magnetic field exhibit three primary flow regimes for fixed aspect ratio and stress free mechanical boundary conditions. The first regime is characterized by 2D dynamics that are steady in time. The second regime is characterized by 3D anisotropic dynamics; in these simulations ohmic dissipation can become comparable to, and in some cases greater than, viscous dissipation. Some of the 3D cases are found to be steady in time, but a transition to unsteady dynamics eventually occurs. Finally, we observe a third regime that is characterized by a large scale mean flow that can become larger in magnitude than the convective flows that feed it.



The first regime is characterized by heat and momentum transport that is more efficient than 3D RBC. Within these 2D states we observe a reduction in the number of convection cells as the Rayleigh number is increased. The heat transport scaling behavior is observed to decrease as the cell number is reduced, though the $Nu \sim Ra^{1/3}$ still describes well the observed behavior. This finding is consistent with the work of Ref.~\citep{gC09}; although there is an optimal aspect ratio for the convective cell that maximizes the overall heat transport, all aspect ratios in steady 2D convection with stress-free boundary conditions follow a $Nu \sim Ra^{1/3}$ scaling provided the Rayleigh number is sufficiently large. The momentum transport in the 2D regime is consistent with the $Re \sim Ra^{2/3}$ discussed in \citeauthor{bW20} \cite{bW20}. 



For sufficiently large Rayleigh number the 2D states will eventually transition to a regime characterized by anisotropic turbulence provided that the Rayleigh number is sufficiently large. Anisotropy persists at the largest Rayleigh numbers investigated even for the smallest value of imposed field strength considered ($Q=10^2$). However, the structure of these anisotropic turbulent states does depend on $Q$; we find that the system prefers a two roll state as $Q$ is increased. The 2D/3D transition occurs when an approximate balance between the inertial and magnetic forces is met, i.e when $Ha/Re=O(1)$. Once again, the $x$-dependence of the 3D state that is first observed ultimately depends on $Q$, with larger scale modulations preferred for the largest values of $Q$.

Our data suggests that heat and momentum transport within the second regime of HMC scale with $Ra$ in a manner that is similar to RBC, i.e.~$Nu \sim Ra^{2/7}$ and $Re \sim Ra^{1/2}$ within our parameter regime. However, for a fixed value of $Ra$ we find that HMC results in less heat transport in comparison to RBC. This reduction in heat transport is not necessarily correlated with reduced momentum transport. Rather, the structure of the convection, characterized by the number of convection rolls, seems to play an important roll in determining whether HMC heat transport is more or less efficient in comparison to 3D RBC. This point might be understood in terms of the theory of Chini \& Cox in which heat transport is maximum when the aspect ratio of a convective cell is order unity. 

The third regime that was observed exhibits a strong mean flow that is directed transverse to the imposed magnetic field. As in studies of convection with a horizontal rotation vector \citep{von2015generation}, we find that the dynamics of this regime are strongly time dependent. Large heat transport is correlated with times in which the mean flow is weak and vice versa. Strong relaxation oscillations are eventually observed, and these flows are computationally prohibitive to simulate given the combination of long time integrations and high spatial resolution that is required to investigate such regimes. Indeed, similar flows (and computational constraints) occur when the imposed field is tilted \citep{jN22}.

The horizontally periodic system employed in the present investigation likely behaves differently than confined systems with vertical boundaries. Indeed, in laboratory experiments of MC with a horizontal magnetic field, the presence of vertical boundaries is known to have a stabilizing influence on convection and the critical Rayleigh number scales as $Ra_c \sim Q^{1/2}$ in the limit $Q\rightarrow \infty$ for electrically insulating boundaries \citep{uB02}. However, the most unstable motions, like the infinite plane layer, consist of rolls aligned with the direction of the imposed magnetic field \citep{uB02,aO03,tY13,yT16,tV18}. The 2D rolls can lead to enhanced heat transport relative to RBC \citep{aO03,tV21}, since the rolls limit horizontal mixing that tends to decrease heat transfer efficiency. A transition from 2D to 3D motions is also observed in confined systems, though the particular parameters that characterize this transition may be quite different than that identified here \citep[e.g.][]{tV21}.


Several extensions of the present study are of interest. It is assumed throughout the entirety of this study that the magnetic Reynolds number $Rm$ is vanishingly small. With magnetically constrained flow configurations, i.e $Q \gg 1$, this approximation is appropriate as the induced currents introduce negligible perturbations to the externally imposed magnetic field. Such an approximation, however, naturally restricts the applicability of this study as the flow field is incapable of dynamo action. Therefore, relaxing the quasi-static approximation would be of interest for understanding how local dynamo action interacts with a large scale horizontal magnetic field. 

For numerical convenience the present investigation was restricted to $Pr=1$, though many types of electrically conducting fluids are characterized by small thermal Prandtl numbers. Liquid metals, as relevant to terrestrial planets and many laboratory experiments, are characterized by $Pr = O(10^{-2})$. Moreover, stellar plasmas can be characterized by Prandtl numbers as small as $Pr = O(10^{-7})$ \citep[e.g.][]{mO03}. It would therefore be of significant interest to extend the present study to smaller values of $Pr$ to determine what influence this parameter has on the transitions and flow regimes for MC with a horizontal magnetic field in the plane periodic geometry. Previous experimental work utilizing liquid metals find strongly time-dependent flow regimes that are likely a consequence of the properties of the working fluid \citep[][]{tY13,yT16,tV18,jY21}. In contrast to liquid metals and plasmas, electrolytic solutions, as used for example in the experiments described by Ref.~\citep{aO03}, are characterized by $Pr > 1$. The results described in the present study might be more applicable to this particular class of electrically conducting fluids.


It would also be of interest to extend the present study by examining the influence of no-slip mechanical boundary conditions. Studies of steady 2D RBC with no-slip boundary conditions show a trend toward the $Nu \sim Ra^{1/3}$ heat transport scaling if the aspect ratio of the convection cell is chosen to optimize heat transport \cite{fW15,bW20}. In contrast to stress-free boundary conditions in which $O(1)$ aspect ratio cells maximize heat transport, studies using no-slip boundary conditions find that small aspect ratio convection cells maximize the heat transport.

\section*{Acknowledgements}
The authors gratefully acknowledge funding from the National Science Foundation through grant EAR-1945270 (MAC). SM acknowledges support from the European Research Council (agreement no. 833848-UEMHP) under the Horizon 2020 programme. 
The Anvil supercomputer at Purdue University was made available through allocation PHY180013 from the Advanced Cyberinfrastructure Coordination Ecosystem: Services \& Support (ACCESS) program, which is supported by National Science Foundation grants \#2138259, \#2138286, \#2138307, \#2137603, and \#2138296. This work also utilized the Alpine high performance computing resource at the University of Colorado Boulder. Alpine is jointly funded by the University of Colorado Boulder, the University of Colorado Anschutz, and Colorado State University.

\pagebreak

\begin{center}
\begin{longtable}{ccccccccc}
\caption{Details of the numerical simulations. The parameters are the Chandrasekhar number ($Q$), Rayleigh number ($Ra$), Nusselt number ($Nu$), Reynolds number ($Re$), time-step size ($\Delta t$), physical space resolution ($N_x \times N_y \times N_z$), average number of convection rolls ($n_r$) and dimensionality of the flow (2D/3D). In all cases the simulation domain size is  $6\lambda_c \times 6 \lambda_c \times 1$, where $\lambda_c \approx 2.83$ is the critical wavelength. \textcolor{black}{The thermal Prandtl number is $Pr=1$ for all simulations.}} \\
$Q$ & $Ra$ & $Nu$ & $Re$ & $\Delta t$ & $N_x \times N_y \times N_z \quad $ & $n_r$ \quad & 2D/3D \\ \hline \hline
\multicolumn{1}{c|}{$10^2$} & $7 \times 10^2$ & 1.12 & 2.41 & $1 \times 10^{-3}$ & $48 \times 144 \times 48$ & 12 & 2D  \\
\multicolumn{1}{c|}{$10^2$} & $1 \times 10^3$ & 1.67 & 7.09 & $1 \times 10^{-3}$ & $48 \times 144 \times 48$ & 10 & 2D  \\
\multicolumn{1}{c|}{$10^2$} & $1.5 \times 10^3$ & 2.04 & 11.4 & $1 \times 10^{-3}$ & $48 \times 144 \times 48$ & 8 & 2D  \\
\multicolumn{1}{c|}{$10^2$} & $2 \times 10^3$ & 2.42 & 15.3 & $1 \times 10^{-3}$ & $48 \times 144 \times 48$ & 8 & 2D \\
\multicolumn{1}{c|}{$10^2$} & $3 \times 10^3$ & 2.51 & 20.1 & $1 \times 10^{-3}$ & $48 \times 192 \times 48$ & 6 & 3D \\
\multicolumn{1}{c|}{$10^2$} & $5 \times 10^3$ & 2.93 & 27.4 & $1 \times 10^{-3}$ & $48 \times 192 \times 48$ & 6 & 3D \\
\multicolumn{1}{c|}{$10^2$} & $7 \times 10^3$ & 3.25 & 33.5 & $1 \times 10^{-3}$ & $96 \times 192 \times 48$ & 6 & 3D \\
\multicolumn{1}{c|}{$10^2$} & $1 \times 10^4$ & 3.57 & 40.1 & $5 \times 10^{-4}$ & $192 \times 192 \times 48$ & 6 & 3D  \\
\multicolumn{1}{c|}{$10^2$} & $1.5 \times 10^4$ & 3.94 & 48.1 & $5 \times 10^{-4}$ & $192 \times 192 \times 48$ & 6 & 3D  \\
\multicolumn{1}{c|}{$10^2$} & $2 \times 10^4$ & 4.23 & 54.3 & $4 \times 10^{-4}$ & $192 \times 288 \times 48$ & 6 & 3D \\
\multicolumn{1}{c|}{$10^2$} & $3 \times 10^4$ & 4.68 & 64.9 & $3 \times 10^{-4}$ & $288 \times 288 \times 48$ & 6 & 3D \\
\multicolumn{1}{c|}{$10^2$} & $5 \times 10^4$ & 5.37 & 81.8 & $2 \times 10^{-4}$ & $384 \times 384 \times 48$ & 6 & 3D \\
\multicolumn{1}{c|}{$10^2$} & $7 \times 10^4$ & 5.88 & 94.8 & $1 \times 10^{-4}$ & $384 \times 384 \times 48$ & 6 & 3D  \\
\multicolumn{1}{c|}{$10^2$} & $1 \times 10^5$ & 6.52 & 111 & $1 \times 10^{-4}$ & $448 \times 448 \times 48$ & 6 & 3D  \\
\multicolumn{1}{c|}{$10^2$} & $1.5 \times 10^5$ & 7.32 & 133 & $5 \times 10^{-5}$ & $448 \times 448 \times 48$ & 6 & 3D  \\
\multicolumn{1}{c|}{$10^2$} & $2 \times 10^5$ & 7.98 & 153 & $5 \times 10^{-5}$ & $576 \times 576 \times 48$ & 6 & 3D  \\
\multicolumn{1}{c|}{$10^2$} & $3 \times 10^5$ & 9.02 & 183 & $4 \times 10^{-5}$ & $576 \times 576 \times 48$ & 6 & 3D  \\
\multicolumn{1}{c|}{$10^2$} & $5 \times 10^5$ & 10.56 & 231 & $4 \times 10^{-5}$ & $768 \times 768 \times 72$ & 6 & 3D  \\
\multicolumn{1}{c|}{$10^2$} & $7 \times 10^5$ & 11.70 & 271 & $7 \times 10^{-6}$ & $900 \times 900 \times 72$ & 6 & 3D  \\
\multicolumn{1}{c|}{$10^2$} & $1 \times 10^6$ & 13.11 & 318 & $5 \times 10^{-7}$ & $900 \times 900 \times 72$ & 6 & 3D  \\ \hline
\multicolumn{1}{c|}{$10^3$} & $7 \times 10^2$ & 1.12 & 2.41 & $1 \times 10^{-3}$ & $48 \times 144 \times 48$ & 12 & 2D  \\
\multicolumn{1}{c|}{$10^3$} & $1 \times 10^3$ & 1.74 & 7.06 & $1 \times 10^{-3}$ & $48 \times 144 \times 48$ & 12 & 2D \\
\multicolumn{1}{c|}{$10^3$} & $1.5 \times 10^3$ & 2.32 & 11.6 & $1 \times 10^{-3}$ & $48 \times 144 \times 48$ & 12 & 2D \\
\multicolumn{1}{c|}{$10^3$} & $2 \times 10^3$ & 2.70 & 15.1 & $1 \times 10^{-3}$ & $48 \times 144 \times 48$ & 12 & 2D \\
\multicolumn{1}{c|}{$10^3$} & $3 \times 10^3$ & 3.22 & 21.05 & $1 \times 10^{-3}$ & $48 \times 192 \times 48$ & 12 & 2D \\ 
\multicolumn{1}{c|}{$10^3$} & $5 \times 10^3$ & 3.92 & 30.9 & $1 \times 10^{-3}$ & $48 \times 192 \times 48$ & 12 & 2D \\ 
\multicolumn{1}{c|}{$10^3$} & $7 \times 10^3$ & 4.06 & 41.2 & $1 \times 10^{-3}$ & $48 \times 288 \times 48$ & 12 & 2D \\ 
\multicolumn{1}{c|}{$10^3$} & $1 \times 10^4$ & 4.63 & 53.2 & $5 \times 10^{-4}$ & $48 \times 288 \times 48$ & 8 & 2D \\ 
\multicolumn{1}{c|}{$10^3$} & $1.5 \times 10^4$ & 5.36 & 70.6 & $5 \times 10^{-4}$ & $48 \times 288 \times 48$ & 8 & 2D \\ 
\multicolumn{1}{c|}{$10^3$} & $2 \times 10^4$ & 5.94 & 86.2 & $3 \times 10^{-4}$ & $48 \times 288 \times 48$ & 8 & 2D \\ 
\multicolumn{1}{c|}{$10^3$} & $3 \times 10^4$ & 4.57 & 88.9 & $2 \times 10^{-4}$ & $48 \times 384 \times 48$ & 4 & 3D \\ 
\multicolumn{1}{c|}{$10^3$} & $5 \times 10^4$ & 4.98 & 104 & $2 \times 10^{-4}$ & $96 \times 384 \times 48$ & 4 & 3D \\ 
\multicolumn{1}{c|}{$10^3$} & $7 \times 10^4$ & 5.26 & 115 & $2 \times 10^{-4}$ & $192 \times 384 \times 48$ & 4 & 3D \\
\multicolumn{1}{c|}{$10^3$} & $1 \times 10^5$ & 5.68 & 131 & $2 \times 10^{-4}$ & $384 \times 384 \times 48$ & 4 & 3D \\
\multicolumn{1}{c|}{$10^3$} & $1.5 \times 10^5$ & 6.41 & 164 & $2 \times 10^{-4}$ & $384 \times 444 \times 48$ & 4 & 3D \\
\multicolumn{1}{c|}{$10^3$} & $2 \times 10^5$ & 6.84 & 183 & $2 \times 10^{-4}$ & $480 \times 480 \times 48$ & 4 & 3D \\
\multicolumn{1}{c|}{$10^3$} & $3 \times 10^5$ & 7.65 & 217 & $2 \times 10^{-4}$ & $576 \times 576 \times 72$ & 4 & 3D \\
\multicolumn{1}{c|}{$10^3$} & $5 \times 10^5$ & 8.69 & 230 & $2 \times 10^{-4}$ & $768 \times 768 \times 72$ & 4 & 3D \\ \hline
\multicolumn{1}{c|}{$10^4$} & $7 \times 10^2$ & 1.12 & 2.41 & $1 \times 10^{-5}$ & $48 \times 144 \times 48$ & 12 & 2D \\
\multicolumn{1}{c|}{$10^4$} & $1 \times 10^3$ & 1.74 & 7.06 & $1 \times 10^{-5}$ & $48 \times 144 \times 48$ & 12 & 2D \\
\multicolumn{1}{c|}{$10^4$} & $1.5 \times 10^3$ & 2.32 & 11.6 & $1 \times 10^{-5}$ & $48 \times 144 \times 48$ & 12 & 2D \\
\multicolumn{1}{c|}{$10^4$} & $2 \times 10^3$ & 2.70 & 15.1 & $1 \times 10^{-5}$ & $48 \times 144 \times 48$ & 12 & 2D \\
\multicolumn{1}{c|}{$10^4$} & $3 \times 10^3$ & 3.22 & 21.1 & $1 \times 10^{-5}$ & $48 \times 192 \times 48$ & 12 & 2D \\
\multicolumn{1}{c|}{$10^4$} & $5 \times 10^3$ & 3.92 & 30.9 & $1 \times 10^{-5}$ & $48 \times 192 \times 48$ & 12 & 2D \\
\multicolumn{1}{c|}{$10^4$} & $7 \times 10^3$ & 4.45 & 39.5 & $1 \times 10^{-5}$ & $48 \times 288 \times 48$ & 12 & 2D \\
\multicolumn{1}{c|}{$10^4$} & $1 \times 10^4$ & 5.09 & 50.9 & $1 \times 10^{-5}$ & $48 \times 288 \times 48$ & 12 & 2D \\
\multicolumn{1}{c|}{$10^4$} & $1.5 \times 10^4$ & 5.91 & 67.5 & $1 \times 10^{-5}$ & $48 \times 288 \times 48$ & 12 & 2D \\
\multicolumn{1}{c|}{$10^4$} & $2 \times 10^4$ & 6.56 & 82.3 & $1 \times 10^{-5}$ & $48 \times 288 \times 48$ & 12 & 2D \\
\multicolumn{1}{c|}{$10^4$} & $3 \times 10^4$ & 7.59 & 109 & $1 \times 10^{-5}$ & $48 \times 384 \times 48$ & 12 & 2D \\
\multicolumn{1}{c|}{$10^4$} & $5 \times 10^4$ & 9.10 & 153 & $1 \times 10^{-5}$ & $48 \times 448 \times 72$ & 12 & 2D \\
\multicolumn{1}{c|}{$10^4$} & $7 \times 10^4$ & 10.25 & 192 & $1 \times 10^{-5}$ & $48 \times 448 \times 72$ & 12 & 2D \\
\multicolumn{1}{c|}{$10^4$} & $1 \times 10^5$ & 11.62 & 244 & $1 \times 10^{-5}$ & $48 \times 448 \times 72$ & 12 & 2D \\
\multicolumn{1}{c|}{$10^4$} & $1.5 \times 10^5$ & 13.38 & 320 & $1 \times 10^{-5}$ & $48 \times 448 \times 72$ & 12 & 2D \\
\multicolumn{1}{c|}{$10^4$} & $2 \times 10^5$ & 12.64 & 368 & $1 \times 10^{-5}$ & $48 \times 576 \times 72$ & 8 & 3D \\
\multicolumn{1}{c|}{$10^4$} & $3 \times 10^5$ & 10.40 & 449 & $1 \times 10^{-5}$ & $96 \times 576 \times 72$ & 4 & 3D \\
\multicolumn{1}{c|}{$10^4$} & $5 \times 10^5$ & 11.28 & 533 & $1 \times 10^{-5}$ & $96 \times 768 \times 72$ & 4 & 3D \\
\multicolumn{1}{c|}{$10^4$} & $7 \times 10^5$ & 11.50 & 558 & $1 \times 10^{-5}$ & $192 \times 768 \times 72$ & 4 & 3D \\
\multicolumn{1}{c|}{$10^4$} & $1 \times 10^6$ & 9.07 & 639 & $1 \times 10^{-5}$ & $288 \times 900 \times 72$ & 2 & 3D \\
\multicolumn{1}{c|}{$10^4$} & $1.5 \times 10^6$ & 10.03 & 719 & $1 \times 10^{-5}$ & $288 \times 900 \times 72$ & 2 & 3D \\
\multicolumn{1}{c|}{$10^4$} & $2 \times 10^6$ & 11.05 & 752 & $5 \times 10^{-6}$ & $384 \times 1024 \times 72$ & 2 & 3D \\
\multicolumn{1}{c|}{$10^4$} & $3 \times 10^6$ & 12.48 & 795 & $5 \times 10^{-6}$ & $384 \times 1024 \times 72$ & 2 & 3D \\ \hline
\multicolumn{1}{c|}{$10^5$} & $7 \times 10^4$ & 10.25 & 192 & $5 \times 10^{-6}$ & $48 \times 448 \times 72$ & 12 & 2D \\
\multicolumn{1}{c|}{$10^5$} & $1 \times 10^5$ & 11.62 & 244 & $5 \times 10^{-6}$ & $48 \times 576 \times 72$ & 12 & 2D \\
\multicolumn{1}{c|}{$10^5$} & $1.5 \times 10^5$ & 13.38 & 320 & $5 \times 10^{-6}$ & $48 \times 768 \times 72$ & 12 & 2D  \\
\multicolumn{1}{c|}{$10^5$} & $2 \times 10^5$ & 14.79 & 388 & $5 \times 10^{-6}$ & $48 \times 768 \times 72$ & 12 & 2D  \\
\multicolumn{1}{c|}{$10^5$} & $3 \times 10^5$ & 17.01 & 508 & $5 \times 10^{-6}$ & $48 \times 768 \times 72$ & 12 & 2D \\ 
\multicolumn{1}{c|}{$10^5$} & $5 \times 10^5$ & 20.27 & 713 & $5 \times 10^{-6}$ & $48 \times 1020 \times 96$ & 12 & 3D \\ 
\multicolumn{1}{c|}{$10^5$} & $7 \times 10^5$ & 14.78 & 873 & $5 \times 10^{-6}$ & $48 \times 1152 \times 96$ & 4 & 3D \\ 
\multicolumn{1}{c|}{$10^5$} & $1 \times 10^6$ & 16.70 & 1108 & $5 \times 10^{-6}$ & $48 \times 1152 \times 96$ & 4 & 3D \\ 
\multicolumn{1}{c|}{$10^5$} & $1.5 \times 10^6$ & 14.57 & 891 & $5 \times 10^{-6}$ & $48 \times 1152 \times 96$ & 4 & 3D \\ 
\multicolumn{1}{c|}{$10^5$} & $2 \times 10^6$ & 16.02 & 1019 & $5 \times 10^{-6}$ & $48 \times 1152 \times 96$ & 4 & 3D \\  \hline
\multicolumn{1}{c|}{$10^6$} & $7 \times 10^2$ & 1.12 & 2.41 & $1 \times 10^{-6}$ & $48 \times 144 \times 48$ & 12 & 2D  \\
\multicolumn{1}{c|}{$10^6$} & $1 \times 10^3$ & 1.74 & 7.06 & $1 \times 10^{-6}$ & $48 \times 192 \times 48$ & 12 & 2D  \\
\multicolumn{1}{c|}{$10^6$} & $2 \times 10^3$ & 2.70 & 14.91 & $1 \times 10^{-6}$ & $48 \times 192 \times 48$ & 12 & 2D  \\
\multicolumn{1}{c|}{$10^6$} & $3 \times 10^3$ & 3.21 & 21.1 & $1 \times 10^{-6}$ & $48 \times 192 \times 48$ & 12 & 2D  \\
\multicolumn{1}{c|}{$10^6$} & $5 \times 10^3$ & 3.92 & 30.9 & $1 \times 10^{-6}$ & $48 \times 192 \times 48$ & 12 & 2D  \\
\multicolumn{1}{c|}{$10^6$} & $7 \times 10^3$ & 4.45 & 39.5 & $1 \times 10^{-6}$ & $48 \times 288 \times 48$ & 12 & 2D  \\
\multicolumn{1}{c|}{$10^6$} & $1 \times 10^4$ & 5.09 & 50.9 & $1 \times 10^{-6}$ & $48 \times 288 \times 48$ & 12 & 2D  \\
\multicolumn{1}{c|}{$10^6$} & $2 \times 10^4$ & 6.56 & 82.2 & $1 \times 10^{-6}$ & $48 \times 384 \times 48$ & 12 & 2D  \\
\multicolumn{1}{c|}{$10^6$} & $3 \times 10^4$ & 7.59 & 109 & $1 \times 10^{-6}$ & $48 \times 384 \times 48$ & 12 & 2D  \\
\multicolumn{1}{c|}{$10^6$} & $5 \times 10^4$ & 9.10 & 153 & $1 \times 10^{-6}$ & $48 \times 444 \times 72$ & 12 & 2D  \\
\multicolumn{1}{c|}{$10^6$} & $7 \times 10^4$ & 10.25 & 192 & $1 \times 10^{-6}$ & $48 \times 444 \times 72$ & 12 & 2D  \\
\multicolumn{1}{c|}{$10^6$} & $1 \times 10^5$ & 11.62 & 244 & $1 \times 10^{-6}$ & $48 \times 444 \times 72$ & 12 & 2D  \\
\multicolumn{1}{c|}{$10^6$} & $2 \times 10^5$ & 14.79 & 388 & $1 \times 10^{-6}$ & $48 \times 768 \times 72$ & 12 & 2D  \\
\multicolumn{1}{c|}{$10^6$} & $3 \times 10^5$ & 17.01 & 508 & $1 \times 10^{-6}$ & $48 \times 768 \times 72$ & 12 & 2D  \\
\multicolumn{1}{c|}{$10^6$} & $5 \times 10^5$ & 20.27 & 713 & $1 \times 10^{-6}$ & $48 \times 900 \times 72$ & 12 & 2D  \\
\multicolumn{1}{c|}{$10^6$} & $7 \times 10^5$ & 22.74 & 891 & $1 \times 10^{-6}$ & $48 \times 1152 \times 96$ & 12 & 2D  \\
\multicolumn{1}{c|}{$10^6$} & $1 \times 10^6$ & 25.69 & 1130 & $1 \times 10^{-6}$ & $48 \times 1152 \times 96$ & 12 & 2D  \\
\multicolumn{1}{c|}{$10^6$} & $1.5 \times 10^6$ & 23.53 & 1541 & $1 \times 10^{-6}$ & $48 \times 1152 \times 96$ & 6 & 2D  \\
\multicolumn{1}{c|}{$10^6$} & $2 \times 10^6$ & 21.16 & 1759 & $1 \times 10^{-6}$ & $48 \times 1152 \times 96$ & 4 & 3D  \\
\multicolumn{1}{c|}{$10^6$} & $3 \times 10^6$ & 24.29 & 2306 & $1 \times 10^{-6}$ & $48 \times 1152 \times 96$ & 4 & 3D  \\
\multicolumn{1}{c|}{$10^6$} & $5 \times 10^6$ & 24.50 & 2316 & $1 \times 10^{-6}$ & $48 \times 1152 \times 96$ & 4 & 3D  \\
\multicolumn{1}{c|}{$10^6$} & $7 \times 10^6$ & 22.47 & 1968 & $1 \times 10^{-6}$ & $48 \times 1296 \times 96$ & 4 & 3D  \\
\multicolumn{1}{c|}{$10^6$} & $1 \times 10^7$ & 23.86 & 4194 & $1 \times 10^{-6}$ & $48 \times 1440 \times 96$ & 2 & 3D  \\
\multicolumn{1}{c|}{$10^6$} & $1.5 \times 10^7$ & 22.09 & 3603 & $1 \times 10^{-6}$ & $192 \times 1440 \times 120$ & 2 & 3D  \\
\multicolumn{1}{c|}{$10^6$} & $2 \times 10^7$ & 23.36 & 4014 & $1 \times 10^{-6}$ & $192 \times 1440 \times 120$ & 2 & 3D  \\  \hline \hline
\label{T:data}
\end{longtable}


\end{center}

\bibliography{References,journal_abbreviations}

\end{document}